\documentclass[aps,  prl, twocolumn, 10pt, superscriptaddress]{revtex4-1}

\pdfoutput=1

\usepackage{lipsum}
\usepackage{graphicx}
\usepackage{amsmath}
\usepackage{amsfonts}

\usepackage{graphicx}

\usepackage{hyperref}
\usepackage{lipsum}
\usepackage{color, soul}

\begin{document}
 \title{Chaotic dynamics in spin-vortex pairs}
 
 \author{A. V. Bondarenko}
 \email{artemb@kth.se}
 \affiliation{Royal Institute of Technology, 10691 Stockholm, Sweden}
 \affiliation{Institute of Magnetism, National Academy of Science, 03142 Kiev, Ukraine}
 
 \author{E. Holmgren}
 \affiliation{Royal Institute of Technology, 10691 Stockholm, Sweden}
 
 \author{Z. W. Li}
 \affiliation{Royal Institute of Technology, 10691 Stockholm, Sweden}
 
 \author{B. A. Ivanov}
 \affiliation{Institute of Magnetism, National Academy of Science, 03142 Kiev, Ukraine}
 \affiliation{National University of Science and Technology ``MISiS'', Moscow, 119049, Russian Federation}
 
 \author{V. Korenivski}
 \affiliation{Royal Institute of Technology, 10691 Stockholm, Sweden}

 \begin{abstract}
  We report on spin-vortex pair dynamics measured at temperatures low enough to suppress stochastic core motion, thereby uncovering the highly 
  non-linear intrinsic dynamics of the system. Our analysis shows that the decoupling of the two vortex cores is resonant and can be enhanced
  by dynamic chaos. We detail the regions of the relevant parameter space, in which the various mechanisms of the resonant core-core dynamics are activated.
  We show that the presence of chaos can reduce the thermally-induced spread in the switching time by up to two orders of magnitude.
 \end{abstract}
 \maketitle
    
 Spin vortices carry a significant fundamental interest\cite{Shinjo930, Wachowiak577, Choe420, PhysRevLett.93.077207, PhysRevLett.97.107204, VanWaeyenberge2006, PhysRevLett.98.087205, PhysRevLett.99.267201, Vansteenkiste2009, Pigeau2010, PhysRevLett.106.197203, Petit-Watelot2012, PhysRevLett.111.247601, PhysRevLett.117.037208} 
 due to their high variety in terms of physical layouts, spin configurations, and types of intra- and
 inter-vortex interactions. Applications of relevance are memory
\cite{VanWaeyenberge2006, Yamada2007, doi:10.1063/1.2998584, Nakano2011, doi:10.1063/1.3551524, Geng201784, doi:10.1063/1.4990990, PhysRevB.70.012404, PhysRevB.74.144419},
 rf signal sources\cite{Pribiag2007,PhysRevLett.100.257201,0022-3727-50-8-085002} as well as biofunctional
 materials\cite{Kim2010,Wong2017}. Vertically stacked vortices, with thin inter-vortex spacers, are rather unique as they can possess a very localized yet very strong core-core
 coupling potential\cite{PhysRevLett.109.097204, PhysRevB.93.054411, Stebliy2017}. Here we study such a tightly spaced vortex pair and focus on its most intriguing configuration, having parallel core polarizations
 and antiparallel vortex chiralities (referred to as the P-AP state; illustrated in Fig. 1a), as the collective dynamics it exhibits are entirely different from
 those of the individual vortices comprising the pair. We show that the system can be made bi-stable with the cores either magnetically and spatially coupled
 (diatomic-molecule type pair) or well separated (‘dissociated molecule’), with highly nonlinear, chaos-enhanced switching dynamics.

 We develop an analytical model based on the Thiele equation\cite{PhysRevLett.30.230} and show that the dynamics of the system can
 be reliably predicted with a set of only two time dependent first order equations -- a bare minimum of dimensions required to have chaotic dynamics. Such simplicity sets our spin vortex pairs apart from other magnetic chaotic systems, studied 
 with either continuous media models\cite{doi:10.1063/1.342378,doi:10.1063/1.340525} or higher-dimensional models\cite{PhysRevB.61.11613}. 
 
 \begin{figure}[t]
  \includegraphics[width=\columnwidth]{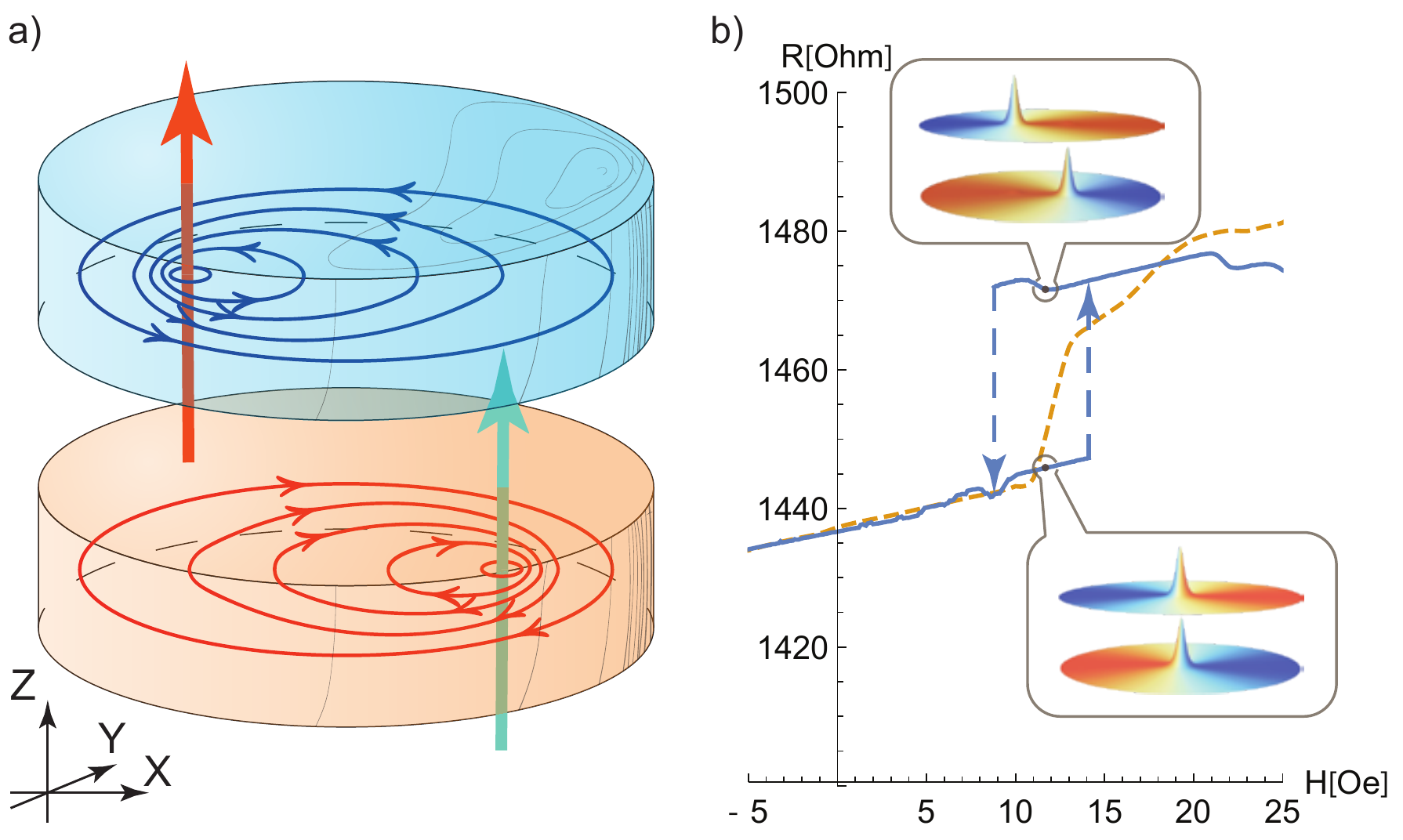}
  \caption{(a) Schematic of the studied vortex pairs, having vertically tightly-spaced magnetic particles, each in a spin vortex state with  
  parallel core polarizations (vertical arrows) and antiparallel chiralities (circular arrows) -- the P-AP vortex pair state. The pair is 
  shown in a \emph{decoupled state}, with a large in-plane core-core separation, as against a \emph{coupled state}, with the two cores on-axis (not shown).
  (b) Measured magnetoresistance of a sample in a P-AP vortex state, with hysteresis at 77~K  (blue) between the coupled and decoupled core-core states
  (illustrated by insets showing corresponding spin maps), and no hysteresis at 300~K due to thermal smearing (dashed orange line).}
  \label{fig:cartoon}
 \end{figure}
 
 The effects of thermal agitation on bi-stable, periodically driven systems are well understood. A dynamically meta-stable state appears as a result of the resonant excitation with its energy elevated, which enhances the stochastic escape rate\cite{doi:10.1063/1.1380368, Devoret1987}.
 Alternatively, when the external force oscillates slower than the system’s characteristic response time, a stochastic resonance can be observed\cite{0305-4470-27-17-001,RevModPhys.70.223}. Thermal effects on chaotic systems were studied to a lesser degree,
 with one example being noise in Josephson junctions\cite{PhysRevLett.55.746, doi:10.1063/1.335642, PhysRevB.53.14546}, important for building voltage standards. Only time-averaged characteristics are of interest in this case since the phase
 of the junction is a periodic variable, making individual switching events non-important.
 
 In this work we investigate the microstates of a spin-vortex pair, including their switching times and trajectories, and show that the presence of dynamic chaos can greatly
 reduce the thermally-induced spread in the switching times, by up to two orders of magnitude. The observed chaos dynamics take place at extremely low excitation
 amplitudes, compared to other reported magnetic systems with chaotic behavior \cite{doi:10.1063/1.340525}.

 \begin{figure}[t]
  \includegraphics[width=\columnwidth]{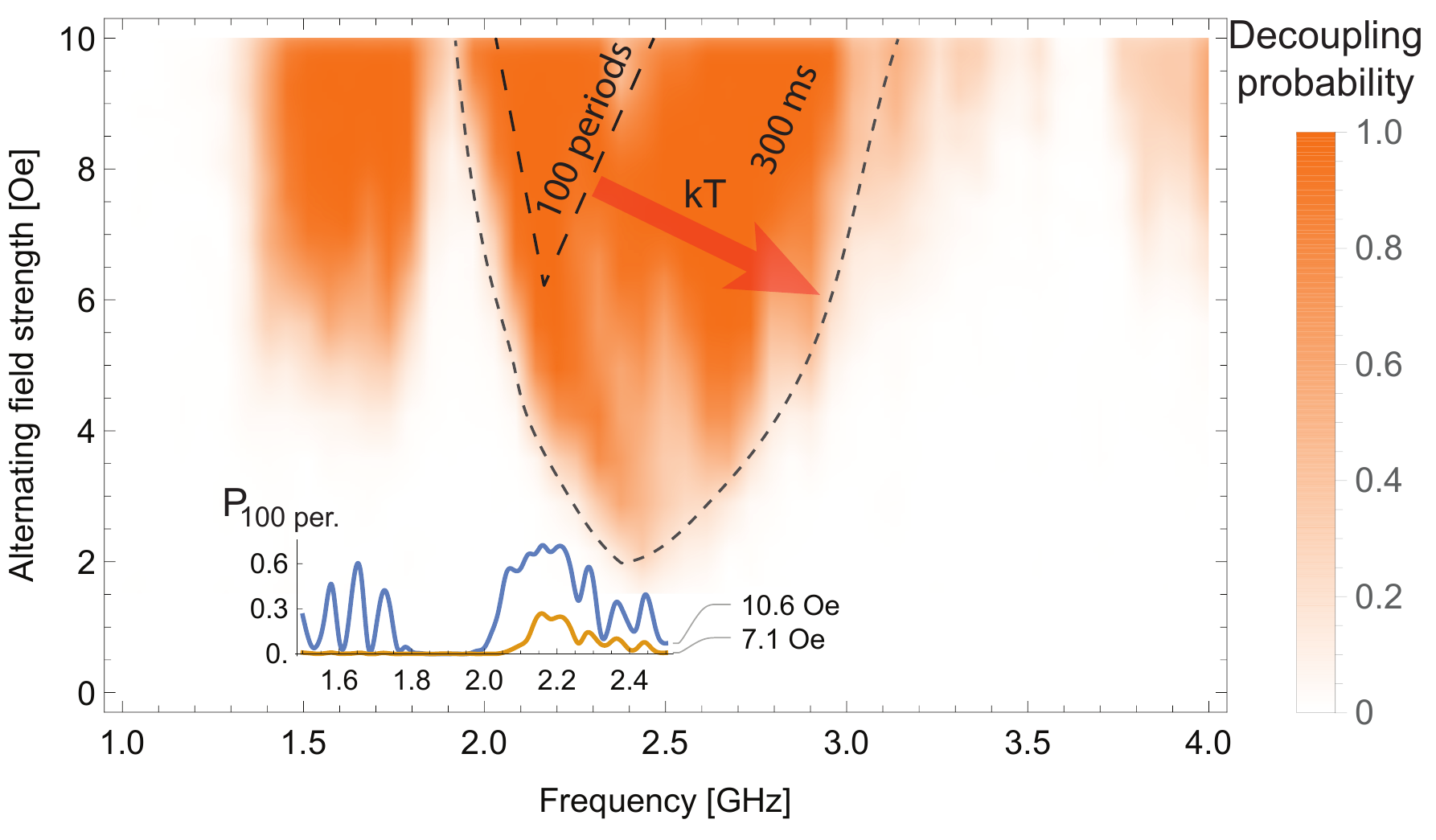}
  \caption{Measured core-decoupling probability as a function of the excitation frequency and field amplitude applied as 300 ms pulse envelopes.
  The inset shows probability-vs-frequency cross-sections for pulse-envelope excitations of 100 periods in duration (40 to 100 ns).}
  \label{fig:map}
 \end{figure}
 

 Our samples were nanopillars containing two vertically stacked elliptical Permalloy (Py) particles, each 350 by 420 nm in-plane and 5 nm thick, separated by a 1 nm thick TaN spacer, integrated with a readout magnetic tunnel junction.
 For measurements, both Py particles were set into a vortex state with parallel core polarizations and antiparallel chiralities (P-AP; Fig.~\ref{fig:cartoon}a), and the magnetic response was measured magnetoresistively
 (see\cite{Gaidis2006,PhysRevB.80.144425}, and Supplementary for experimental details).

 Figure~\ref{fig:cartoon}b shows the magnetoresistance of a typical sample in the P-AP vortex state at 77~K and room temperature (RT). The well-defined $R-H$ hysteresis
 observed at 77~K at about 10-15~Oe is due to decoupling and recoupling of the two vortex cores, and is smeared out at RT. With all measurements done at 77~K and the dc field kept at mid-hysteresis, we focus below on this key hysteretic transition in the system
 between its coupled and decoupled states (with tightly-bound and dissociated cores, as illustrated by the micromagnetically simulated spin maps for the two states, discussed in more detail in Supplementary), which exhibits unique, chaos-enhanced dynamics.
 
 \begin{figure*}[t]
  \includegraphics[width=\textwidth]{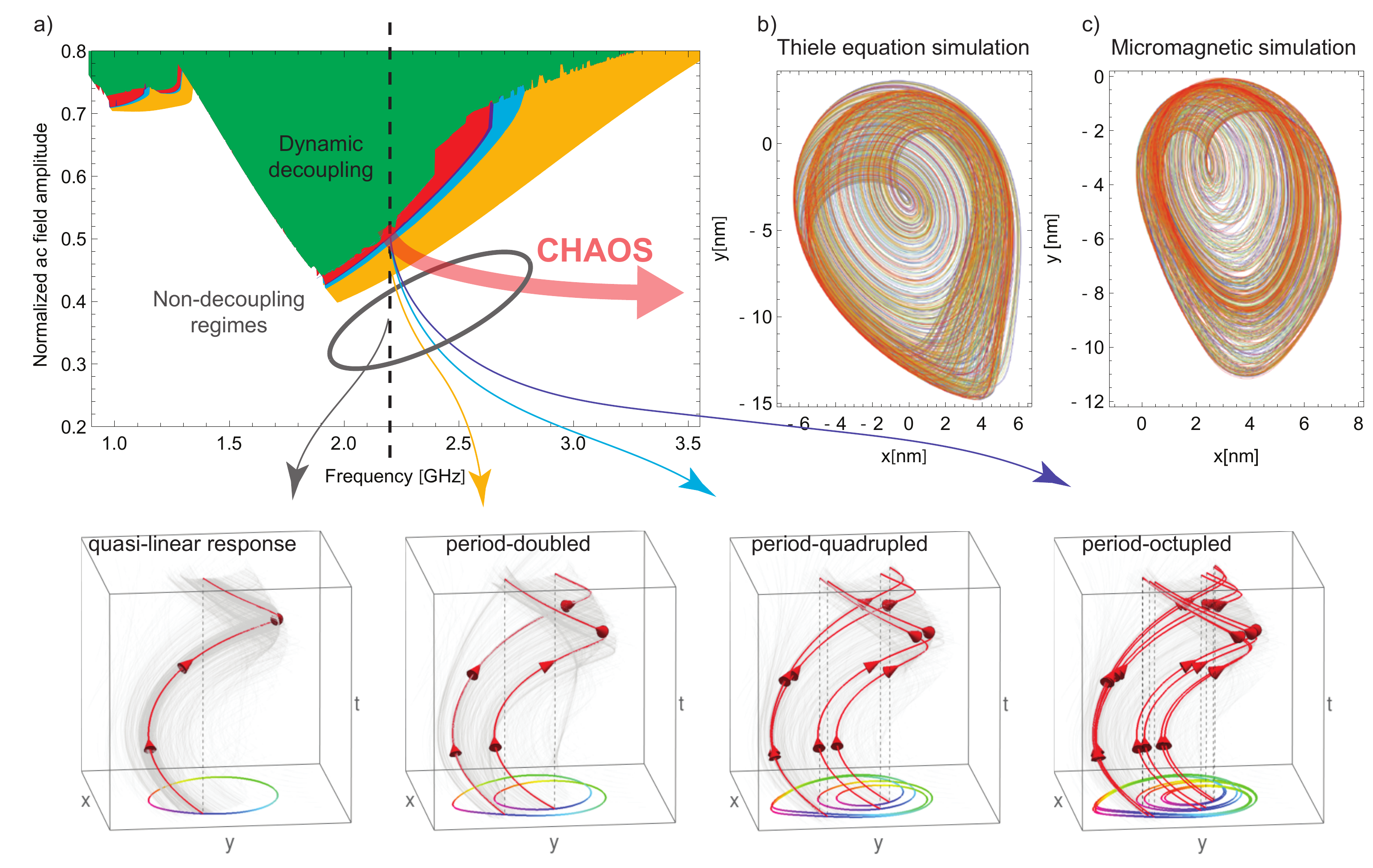}
  \caption{(a) Bifurcation map of a P-AP vortex pair, with the bottom panels showing the progression of qualitative changes in the core trajectories, 
  from linear to non-linear with sequential period-doubling and eventually chaotic. (b) Example of a set of chaotic core trajectories obtained using
  our analytical model. (c) A micromagnetic simulation of the same configuration as used in (b), with the comparison fully validating the analytical
  approach employed.}
  \label{fig:theorygroup}
 \end{figure*}
 
 The core-core decoupling process was mapped out versus frequency and amplitude of the field excitation applied as a pulse envelope of 300 ms in duration.
 Such a pulse envelope of a given amplitude and frequency, with a subsequent readout of the junction state as to decoupled/coupled,
 was repeated 35 times to yield the core-core switching probability for a given point in the parameter space, shown in Fig.~\ref{fig:map}. The bulk of
 the switching region (orange) is somewhat lower in frequency than the small-signal rotational resonance for the system 
 (about 3~GHz\cite{PhysRevLett.109.097204}, see Suppl.), expected since the dc bias field corresponding to mid-hysteresis used in the measurement increases
 the core-core separation of the coupled state, thereby lowering the rotational  frequency of the pair. At higher excitation amplitudes the switching 
 probability map shows a complex structure with a sub-band centered at 1.5 GHz.

 The picture changes for shorter pulse envelopes. 100 period pulses (about 40 ns, versus 300 ms for the main $P(A,f)$ phase space of Fig.~\ref{fig:map}),
 timed to take into consideration the geometry-modified effective damping constant\cite{doi:10.1063/1.2221904} of 0.1, are sufficient to establish a steady-state oscillation while short enough to suppress
 thermal escape events at 77~K. The corresponding response, shown in the inset to Fig.~\ref{fig:map}, is qualitatively the same as the main probability-vs-frequency
 peak, but requires a significantly higher amplitude (illustrated by the dashed triangle in the 300~ms data). The main switching map (300~ms)
 is broadened toward higher frequencies, along the ‘$kT$-arrow’. This is to be expected since for a given field amplitude the high-frequency
 forced oscillations are located deeper in the potential well, where a longer time is needed to encounter a suitable thermal excitation event.
 Thus, by varying the excitation duration and amplitude we can study the various regimes of the vortex-pair dynamics with its rich phase
 space -- essentially deterministic, stochastic, weakly or highly nonlinear, as well as chaotic.


 We use the Thiele equation framework established earlier\cite{PhysRevLett.109.097204} to describe the vortex motion in the presence of 
 a strong core-core interaction. In this model, the core-decoupling dynamics are fully described
 by using only the separation between the cores, $\mathbf{x}=\mathbf{X}_1-\mathbf{X}_2$, where $\mathbf{X}_{1,2}$ are the in-plane coordinates of the two cores.
 The collective motion of the pair described by $\mathbf{X}_1+\mathbf{X}_2$ can be disregarded for the discussion herein since the intermode coupling is 
 negligible, deprecated further by the immense difference of the respective characteristic frequencies (of the core-core rotational motion versus that of the pair's center).

 The resulting equations of motion for the separation vector are
 \begin{equation}
  \left[\mathbf{e}_z\times\dot{\mathbf{x}}\right]=\omega(|\mathbf{x}|)\mathbf{x}+\lambda\dot{\mathbf{x}}+C\left(\mathbf{H}_{bias}+\mathbf{h}(t)\right)
  \label{eq:model},
 \end{equation}
 where $\omega=\frac{\partial U(x|X=0)/\partial x}{2Gx}$ is the intrinsic oscillation frequency dependent upon the oscillation amplitude $x$, $\lambda=\pi\alpha\ln(R/\Delta)$\cite{doi:10.1063/1.2221904, 0295-5075-103-5-57004}, with the micromagnetic
 damping constant $\alpha$, $R$ -- radius of the particle, $\Delta$ -- vortex core size, $\mathbf{h}$ and $\mathbf{H}_{bias}$ -- ac and dc magnetic fields. The form of the core-core potential, $U$, was discussed in detail in\cite{PhysRevLett.109.097204}. Under an external time-dependent force the phase space 
 of the system becomes three dimensional, since now the motion is determined not only by the starting position but also by the starting time. Further details  of the model are  discussed in Supplementary.

 We illustrate the above model by numerically plotting in Fig.~\ref{fig:theorygroup}a the bifurcation map, which displays the different dynamic regimes in 
 our system. Here and below, the ac field amplitude is in Oersteds rms normalized to the biasing field magnitude. One can see that the main core-decoupling
 map is centered at 2.0~GHz. Additionally, a lower-frequency switching sub-band, itself with sub-structure, is visible at around 1.5~GHz. Comparing
 Fig.~\ref{fig:theorygroup} with Fig.~\ref{fig:map}, one can see that the theoretical and experimental core-core decoupling maps are in
 good agreement in terms of the general layout as well as the sub-structure.

 The right wing of the core-switching map, its high-frequency side, is particularly interesting. Here, the system undergoes a period-doubling cascade 
 (yellow, blue, and purple regions in Fig.~\ref{fig:theorygroup}a), giving rise to chaotic dynamics (red) right at the edge of the dynamic-decoupling regime (green). 
 In the chaotic regime, the core trajectories are never repeated and do not settle into a steady state (Fig.~\ref{fig:theorygroup}b,c).
 
 The period-doubling can be qualitatively explained as follows: at some amplitude the applied ac field becomes strong enough to pull the cores from the
 bottom of the coupled-potential well up toward the edge where the intrinsic oscillation frequency of the pair is lower and the cores detune from the
 external excitation. The cores then fall back into the well under the influence of magnetic friction and the now out-of-phase ac field, and after one
 period of recovery the cycle is repeated. The period-doubled response can thus be split into a slow
 close-to-the-edge motion with a high chance of escape, and a fast recovery motion at the bottom of the well, with the two strictly alternating.
 At still higher excitation amplitudes, in the chaotic regime, the core trajectories can cross the free motion separatrix, defined such that if the field was turned
 off at that exact point the cores would not return to the coupled-state well and a switching would occur.

 The results of the analytical model were compared to micromagnetic simulations performed using the Mumax3[33] package. The resulting chaotic
 trajectories are compared in Fig. 3c versus Fig. 3b. We observe that the micromagnetic chaotic trajectory shares the same shape and qualitative evolution
 as the analytical one obtained using (1), which is a strong validation for the model used. 

 \begin{figure*}[t]
  \center\includegraphics[width=\textwidth]{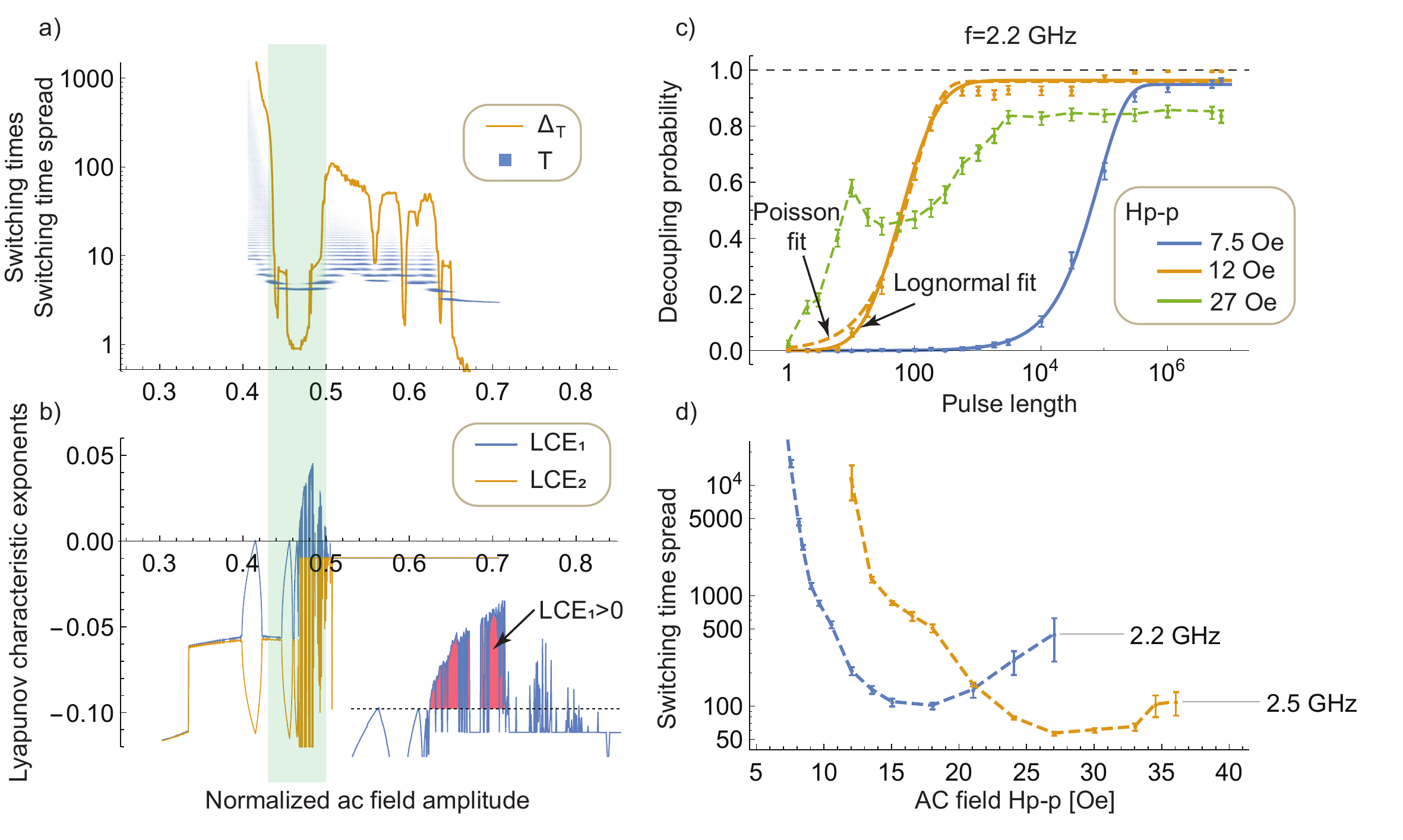}
  \caption{(a) Simulated decoupling time, $\textnormal{T}$, and decoupling time spread, $\Delta_\textnormal{T}$,
  as a function of the ac field amplitude of fixed frequency 2.1~GHz; the drastic reduction in $\textnormal{T}$ and $\Delta_\textnormal{T}$ at about h=0.46 are the chaos-induced reduction
  in the core-switching time and the width of the switching transition. (b) Simulated Lyapunov characteristic exponents; the inset
  shows the ‘pinching’ section of LCE into the positive range, reflecting the chaotic character of the core motion.  (c) Measured 
  core-switching probability as a function of the length of the ac 
  field pulse envelope, expressed in units of one period. (d) Measured width of the core-switching transition
  (in Fig. 4c) versus the applied field amplitude for 2.2 and 2.5 GHz; to be directly compared with the results of the analytical model in Fig. 4b. 
  All measurements performed at 77 K, and mid-hysteresis biasing dc field of 17~Oe. }
  \label{fig:graphs}
 \end{figure*}
 
 In order to understand role of thermal fluctuations we next discuss the short-pulse dynamics. We use the fluctuation dissipation theorem to determine the
 random force dispersion $\Gamma$ [Suppl]. We then use the  stochastic Runge-Kutta numerical algorithm to obtain the distribution of switching 
 times in our system. The representative distributions of the switching time (T) as well as its spread ($\Delta_\textnormal{T}$) are shown 
 in Fig.~\ref{fig:graphs}a versus the ac field amplitude. For a typical thermally-agitated system, the transition width decreases monotonously
 since the relative role of the stochastic effects decreases with increasing the amplitude of the external force. The distribution is, however,
 highly non-monotonous, with distinct minima of one to two orders of magnitude, superposed on to a gradual decay.  

 The first, most pronounced minimum precisely coincides with the amplitude range where the dynamics becomes chaotic, as evidenced by the corresponding
 Lyapunov characteristic exponents (non-zero $\textnormal{LCE}_{1,2}$) shown in Fig.~4b, with one of them crossing into the positive range, indicating the onset of chaos. The close proximity of
 the chaotic trajectories to the decoupled state's basin of attraction, combined with the fractal nature of the chaotic attractor, make the energy
 barrier to switching arbitrarily small [Supplementary Fig.~S5]. At the same time the two attractors posses different degrees of stability, such that at 
 low temperatures the cores, once they decouple, cannot be efficiently recaptured by the chaos attractor because of the essentially absolute stability 
 of the core trajectories in the decoupled basin.

 We have observed this theoretically predicted chaos signature in a direct experiment. The switching time and its statistical distribution 
 were measured by varying the pulse envelope of the applied ac field of given amplitude and frequency from 1 to $10^7$ cycles and recording
 whether the vortex pair switched into the decoupled state, with the entire sequence repeated 1000 times to obtain accurate statistics.
 Three distinct dynamic regimes are observed, shown in Fig.~\ref{fig:graphs}c, as the amplitude is increased.

 The switching probability for low fields (blue curve in Fig.~\ref{fig:graphs}c) is well described by the Poisson distribution, 
 with the switching rates limited by the rate of thermal fluctuations with energy sufficient for lifting the cores out of the 
 coupled attractor. The rate of such fluctuations is lower than the characteristic relaxation time of the system for the given parameters,
 which allows the cores to relax to dynamically stable trajectories between the thermal-escape events.

 At higher amplitudes (orange curve) the dynamically stable trajectories, on average, increase in radius and pass closer to the separatrix,
 which makes lower-energy thermal fluctuations sufficient for activating core-decoupling. At the same time, the number of fluctuations
 with energy comparable to the switching threshold is much larger and their effects can multiply within the relaxation time for a given
 trajectory. We point out, that, as expected, the Poisson distribution no longer is accurate in the limit where the intrinsic dynamics
 of the system is dominating and, as a result, a log-normal distribution provides a much better fit to the experimental data (as shown in Fig.~\ref{fig:graphs}c). 

 Still higher amplitudes (green curve in Fig.~\ref{fig:graphs}c) alter the probability in a qualitative way, such that it does not saturate at unity due to
 significant recoupling. The recoupling probability is high due to the forced high-energy oscillations within the decoupled well post-switching,
 which in turn can undergo thermal excitation events, bringing the system back into the coupled attractor.

 The spread in the switching time (the width of the switching transition) extracted from the switching probability using the above Poissonian and log-normal fitting,
 is shown in Fig.~\ref{fig:graphs}d for two frequencies near the core-core resonance. The observed non-monotonous behavior is in excellent agreement with the theoretical prediction,
 with the transition width going through a deep minimum at intermediate field amplitudes. The transition width is not straightforward to define at the highest
 amplitudes, which results in the data cutoff in Fig.~\ref{fig:graphs}d at 27 and 36 Oe for 2.2 GHz and 2.5 GHz, respectively. The shift of the distribution to lower
 field amplitudes with lowering the frequency as well as the corresponding decrease in the depth of the $\Delta_\textnormal{T}$-vs-h minimum in Fig.~\ref{fig:graphs}d are consistent
 with the changes expected theoretically (not shown) as one moves along the right wing of the core-decoupling bifurcation map of Fig.~\ref{fig:theorygroup}a.
 
 A vortex pair with a hysteretic core-core bi-stability is used to study dynamic chaos in a nanoscale spin system. 
 The observed core-core switching is chaos-enhanced by up to two orders of magnitude in speed and can take place at ultra-low resonant fields.
 These results expand the knowledge base of nanomagnetism, demonstrating a system with performance benefiting from dynamic chaos, of relevance for applications 
 in spintronics. The uncovered details of the core-core dissociation can serve as a model for other, nonmagnetic atomic, molecular, and nanosystems.
 
 Acknowledgements: Support from the Swedish Research Council (VR Grant No. 2014-4548), the National Academy of Sciences of Ukraine
 via project number 1/17-N and by the Ministry of Education and Science of the Russian Federation in the framework of Increase Competitiveness
 Program of  NUST «MISiS» (K2-2017-005), implemented by a governmental decree dated 16th of March 2013, N 211 are gratefully acknowledged.

 \bibliographystyle{apsrev4-1}
 \bibliography{citations}

\begin{thebibliography}{45}%
\makeatletter
\providecommand \@ifxundefined [1]{%
 \@ifx{#1\undefined}
}%
\providecommand \@ifnum [1]{%
 \ifnum #1\expandafter \@firstoftwo
 \else \expandafter \@secondoftwo
 \fi
}%
\providecommand \@ifx [1]{%
 \ifx #1\expandafter \@firstoftwo
 \else \expandafter \@secondoftwo
 \fi
}%
\providecommand \natexlab [1]{#1}%
\providecommand \enquote  [1]{``#1''}%
\providecommand \bibnamefont  [1]{#1}%
\providecommand \bibfnamefont [1]{#1}%
\providecommand \citenamefont [1]{#1}%
\providecommand \href@noop [0]{\@secondoftwo}%
\providecommand \href [0]{\begingroup \@sanitize@url \@href}%
\providecommand \@href[1]{\@@startlink{#1}\@@href}%
\providecommand \@@href[1]{\endgroup#1\@@endlink}%
\providecommand \@sanitize@url [0]{\catcode `\\12\catcode `\$12\catcode
  `\&12\catcode `\#12\catcode `\^12\catcode `\_12\catcode `\%12\relax}%
\providecommand \@@startlink[1]{}%
\providecommand \@@endlink[0]{}%
\providecommand \url  [0]{\begingroup\@sanitize@url \@url }%
\providecommand \@url [1]{\endgroup\@href {#1}{\urlprefix }}%
\providecommand \urlprefix  [0]{URL }%
\providecommand \Eprint [0]{\href }%
\providecommand \doibase [0]{http://dx.doi.org/}%
\providecommand \selectlanguage [0]{\@gobble}%
\providecommand \bibinfo  [0]{\@secondoftwo}%
\providecommand \bibfield  [0]{\@secondoftwo}%
\providecommand \translation [1]{[#1]}%
\providecommand \BibitemOpen [0]{}%
\providecommand \bibitemStop [0]{}%
\providecommand \bibitemNoStop [0]{.\EOS\space}%
\providecommand \EOS [0]{\spacefactor3000\relax}%
\providecommand \BibitemShut  [1]{\csname bibitem#1\endcsname}%
\let\auto@bib@innerbib\@empty
\bibitem [{\citenamefont {Shinjo}\ \emph {et~al.}(2000)\citenamefont {Shinjo},
  \citenamefont {Okuno}, \citenamefont {Hassdorf}, \citenamefont {Shigeto},\
  and\ \citenamefont {Ono}}]{Shinjo930}%
  \BibitemOpen
  \bibfield  {author} {\bibinfo {author} {\bibfnamefont {T.}~\bibnamefont
  {Shinjo}}, \bibinfo {author} {\bibfnamefont {T.}~\bibnamefont {Okuno}},
  \bibinfo {author} {\bibfnamefont {R.}~\bibnamefont {Hassdorf}}, \bibinfo
  {author} {\bibfnamefont {K.}~\bibnamefont {Shigeto}}, \ and\ \bibinfo
  {author} {\bibfnamefont {T.}~\bibnamefont {Ono}},\ }\href {\doibase
  10.1126/science.289.5481.930} {\bibfield  {journal} {\bibinfo  {journal}
  {Science}\ }\textbf {\bibinfo {volume} {289}},\ \bibinfo {pages} {930}
  (\bibinfo {year} {2000})},\ \Eprint
  {http://arxiv.org/abs/http://science.sciencemag.org/content/289/5481/930.full.pdf}
  {http://science.sciencemag.org/content/289/5481/930.full.pdf} \BibitemShut
  {NoStop}%
\bibitem [{\citenamefont {Wachowiak}\ \emph {et~al.}(2002)\citenamefont
  {Wachowiak}, \citenamefont {Wiebe}, \citenamefont {Bode}, \citenamefont
  {Pietzsch}, \citenamefont {Morgenstern},\ and\ \citenamefont
  {Wiesendanger}}]{Wachowiak577}%
  \BibitemOpen
  \bibfield  {author} {\bibinfo {author} {\bibfnamefont {A.}~\bibnamefont
  {Wachowiak}}, \bibinfo {author} {\bibfnamefont {J.}~\bibnamefont {Wiebe}},
  \bibinfo {author} {\bibfnamefont {M.}~\bibnamefont {Bode}}, \bibinfo {author}
  {\bibfnamefont {O.}~\bibnamefont {Pietzsch}}, \bibinfo {author}
  {\bibfnamefont {M.}~\bibnamefont {Morgenstern}}, \ and\ \bibinfo {author}
  {\bibfnamefont {R.}~\bibnamefont {Wiesendanger}},\ }\href {\doibase
  10.1126/science.1075302} {\bibfield  {journal} {\bibinfo  {journal}
  {Science}\ }\textbf {\bibinfo {volume} {298}},\ \bibinfo {pages} {577}
  (\bibinfo {year} {2002})},\ \Eprint
  {http://arxiv.org/abs/http://science.sciencemag.org/content/298/5593/577.full.pdf}
  {http://science.sciencemag.org/content/298/5593/577.full.pdf} \BibitemShut
  {NoStop}%
\bibitem [{\citenamefont {Choe}\ \emph {et~al.}(2004)\citenamefont {Choe},
  \citenamefont {Acremann}, \citenamefont {Scholl}, \citenamefont {Bauer},
  \citenamefont {Doran}, \citenamefont {St{\"o}hr},\ and\ \citenamefont
  {Padmore}}]{Choe420}%
  \BibitemOpen
  \bibfield  {author} {\bibinfo {author} {\bibfnamefont {S.-B.}\ \bibnamefont
  {Choe}}, \bibinfo {author} {\bibfnamefont {Y.}~\bibnamefont {Acremann}},
  \bibinfo {author} {\bibfnamefont {A.}~\bibnamefont {Scholl}}, \bibinfo
  {author} {\bibfnamefont {A.}~\bibnamefont {Bauer}}, \bibinfo {author}
  {\bibfnamefont {A.}~\bibnamefont {Doran}}, \bibinfo {author} {\bibfnamefont
  {J.}~\bibnamefont {St{\"o}hr}}, \ and\ \bibinfo {author} {\bibfnamefont
  {H.~A.}\ \bibnamefont {Padmore}},\ }\href {\doibase 10.1126/science.1095068}
  {\bibfield  {journal} {\bibinfo  {journal} {Science}\ }\textbf {\bibinfo
  {volume} {304}},\ \bibinfo {pages} {420} (\bibinfo {year} {2004})},\ \Eprint
  {http://arxiv.org/abs/http://science.sciencemag.org/content/304/5669/420.full.pdf}
  {http://science.sciencemag.org/content/304/5669/420.full.pdf} \BibitemShut
  {NoStop}%
\bibitem [{\citenamefont {Buess}\ \emph {et~al.}(2004)\citenamefont {Buess},
  \citenamefont {H\"ollinger}, \citenamefont {Haug}, \citenamefont
  {Perzlmaier}, \citenamefont {Krey}, \citenamefont {Pescia}, \citenamefont
  {Scheinfein}, \citenamefont {Weiss},\ and\ \citenamefont
  {Back}}]{PhysRevLett.93.077207}%
  \BibitemOpen
  \bibfield  {author} {\bibinfo {author} {\bibfnamefont {M.}~\bibnamefont
  {Buess}}, \bibinfo {author} {\bibfnamefont {R.}~\bibnamefont {H\"ollinger}},
  \bibinfo {author} {\bibfnamefont {T.}~\bibnamefont {Haug}}, \bibinfo {author}
  {\bibfnamefont {K.}~\bibnamefont {Perzlmaier}}, \bibinfo {author}
  {\bibfnamefont {U.}~\bibnamefont {Krey}}, \bibinfo {author} {\bibfnamefont
  {D.}~\bibnamefont {Pescia}}, \bibinfo {author} {\bibfnamefont {M.~R.}\
  \bibnamefont {Scheinfein}}, \bibinfo {author} {\bibfnamefont
  {D.}~\bibnamefont {Weiss}}, \ and\ \bibinfo {author} {\bibfnamefont {C.~H.}\
  \bibnamefont {Back}},\ }\href {\doibase 10.1103/PhysRevLett.93.077207}
  {\bibfield  {journal} {\bibinfo  {journal} {Phys. Rev. Lett.}\ }\textbf
  {\bibinfo {volume} {93}},\ \bibinfo {pages} {077207} (\bibinfo {year}
  {2004})}\BibitemShut {NoStop}%
\bibitem [{\citenamefont {Kasai}\ \emph {et~al.}(2006)\citenamefont {Kasai},
  \citenamefont {Nakatani}, \citenamefont {Kobayashi}, \citenamefont {Kohno},\
  and\ \citenamefont {Ono}}]{PhysRevLett.97.107204}%
  \BibitemOpen
  \bibfield  {author} {\bibinfo {author} {\bibfnamefont {S.}~\bibnamefont
  {Kasai}}, \bibinfo {author} {\bibfnamefont {Y.}~\bibnamefont {Nakatani}},
  \bibinfo {author} {\bibfnamefont {K.}~\bibnamefont {Kobayashi}}, \bibinfo
  {author} {\bibfnamefont {H.}~\bibnamefont {Kohno}}, \ and\ \bibinfo {author}
  {\bibfnamefont {T.}~\bibnamefont {Ono}},\ }\href {\doibase
  10.1103/PhysRevLett.97.107204} {\bibfield  {journal} {\bibinfo  {journal}
  {Phys. Rev. Lett.}\ }\textbf {\bibinfo {volume} {97}},\ \bibinfo {pages}
  {107204} (\bibinfo {year} {2006})}\BibitemShut {NoStop}%
\bibitem [{\citenamefont {Van~Waeyenberge}\ \emph {et~al.}(2006)\citenamefont
  {Van~Waeyenberge}, \citenamefont {Puzic}, \citenamefont {Stoll},
  \citenamefont {Chou}, \citenamefont {Tyliszczak}, \citenamefont {Hertel},
  \citenamefont {F{\"a}hnle}, \citenamefont {Br{\"u}ckl}, \citenamefont {Rott},
  \citenamefont {Reiss}, \citenamefont {Neudecker}, \citenamefont {Weiss},
  \citenamefont {Back},\ and\ \citenamefont {Sch{\"u}tz}}]{VanWaeyenberge2006}%
  \BibitemOpen
  \bibfield  {author} {\bibinfo {author} {\bibfnamefont {B.}~\bibnamefont
  {Van~Waeyenberge}}, \bibinfo {author} {\bibfnamefont {A.}~\bibnamefont
  {Puzic}}, \bibinfo {author} {\bibfnamefont {H.}~\bibnamefont {Stoll}},
  \bibinfo {author} {\bibfnamefont {K.~W.}\ \bibnamefont {Chou}}, \bibinfo
  {author} {\bibfnamefont {T.}~\bibnamefont {Tyliszczak}}, \bibinfo {author}
  {\bibfnamefont {R.}~\bibnamefont {Hertel}}, \bibinfo {author} {\bibfnamefont
  {M.}~\bibnamefont {F{\"a}hnle}}, \bibinfo {author} {\bibfnamefont
  {H.}~\bibnamefont {Br{\"u}ckl}}, \bibinfo {author} {\bibfnamefont
  {K.}~\bibnamefont {Rott}}, \bibinfo {author} {\bibfnamefont {G.}~\bibnamefont
  {Reiss}}, \bibinfo {author} {\bibfnamefont {I.}~\bibnamefont {Neudecker}},
  \bibinfo {author} {\bibfnamefont {D.}~\bibnamefont {Weiss}}, \bibinfo
  {author} {\bibfnamefont {C.~H.}\ \bibnamefont {Back}}, \ and\ \bibinfo
  {author} {\bibfnamefont {G.}~\bibnamefont {Sch{\"u}tz}},\ }\href
  {http://dx.doi.org/10.1038/nature05240} {\bibfield  {journal} {\bibinfo
  {journal} {Nature}\ }\textbf {\bibinfo {volume} {444}},\ \bibinfo {pages}
  {461 EP } (\bibinfo {year} {2006})}\BibitemShut {NoStop}%
\bibitem [{\citenamefont {Choi}\ \emph {et~al.}(2007)\citenamefont {Choi},
  \citenamefont {Lee}, \citenamefont {Guslienko},\ and\ \citenamefont
  {Kim}}]{PhysRevLett.98.087205}%
  \BibitemOpen
  \bibfield  {author} {\bibinfo {author} {\bibfnamefont {S.}~\bibnamefont
  {Choi}}, \bibinfo {author} {\bibfnamefont {K.-S.}\ \bibnamefont {Lee}},
  \bibinfo {author} {\bibfnamefont {K.~Y.}\ \bibnamefont {Guslienko}}, \ and\
  \bibinfo {author} {\bibfnamefont {S.-K.}\ \bibnamefont {Kim}},\ }\href
  {\doibase 10.1103/PhysRevLett.98.087205} {\bibfield  {journal} {\bibinfo
  {journal} {Phys. Rev. Lett.}\ }\textbf {\bibinfo {volume} {98}},\ \bibinfo
  {pages} {087205} (\bibinfo {year} {2007})}\BibitemShut {NoStop}%
\bibitem [{\citenamefont {Buchanan}\ \emph {et~al.}(2007)\citenamefont
  {Buchanan}, \citenamefont {Grimsditch}, \citenamefont {Fradin}, \citenamefont
  {Bader},\ and\ \citenamefont {Novosad}}]{PhysRevLett.99.267201}%
  \BibitemOpen
  \bibfield  {author} {\bibinfo {author} {\bibfnamefont {K.~S.}\ \bibnamefont
  {Buchanan}}, \bibinfo {author} {\bibfnamefont {M.}~\bibnamefont
  {Grimsditch}}, \bibinfo {author} {\bibfnamefont {F.~Y.}\ \bibnamefont
  {Fradin}}, \bibinfo {author} {\bibfnamefont {S.~D.}\ \bibnamefont {Bader}}, \
  and\ \bibinfo {author} {\bibfnamefont {V.}~\bibnamefont {Novosad}},\ }\href
  {\doibase 10.1103/PhysRevLett.99.267201} {\bibfield  {journal} {\bibinfo
  {journal} {Phys. Rev. Lett.}\ }\textbf {\bibinfo {volume} {99}},\ \bibinfo
  {pages} {267201} (\bibinfo {year} {2007})}\BibitemShut {NoStop}%
\bibitem [{\citenamefont {Vansteenkiste}\ \emph {et~al.}(2009)\citenamefont
  {Vansteenkiste}, \citenamefont {Chou}, \citenamefont {Weigand}, \citenamefont
  {Curcic}, \citenamefont {Sackmann}, \citenamefont {Stoll}, \citenamefont
  {Tyliszczak}, \citenamefont {Woltersdorf}, \citenamefont {Back},
  \citenamefont {Sch{\"u}tz},\ and\ \citenamefont
  {Van~Waeyenberge}}]{Vansteenkiste2009}%
  \BibitemOpen
  \bibfield  {author} {\bibinfo {author} {\bibfnamefont {A.}~\bibnamefont
  {Vansteenkiste}}, \bibinfo {author} {\bibfnamefont {K.~W.}\ \bibnamefont
  {Chou}}, \bibinfo {author} {\bibfnamefont {M.}~\bibnamefont {Weigand}},
  \bibinfo {author} {\bibfnamefont {M.}~\bibnamefont {Curcic}}, \bibinfo
  {author} {\bibfnamefont {V.}~\bibnamefont {Sackmann}}, \bibinfo {author}
  {\bibfnamefont {H.}~\bibnamefont {Stoll}}, \bibinfo {author} {\bibfnamefont
  {T.}~\bibnamefont {Tyliszczak}}, \bibinfo {author} {\bibfnamefont
  {G.}~\bibnamefont {Woltersdorf}}, \bibinfo {author} {\bibfnamefont {C.~H.}\
  \bibnamefont {Back}}, \bibinfo {author} {\bibfnamefont {G.}~\bibnamefont
  {Sch{\"u}tz}}, \ and\ \bibinfo {author} {\bibfnamefont {B.}~\bibnamefont
  {Van~Waeyenberge}},\ }\href {http://dx.doi.org/10.1038/nphys1231} {\bibfield
  {journal} {\bibinfo  {journal} {Nature Physics}\ }\textbf {\bibinfo {volume}
  {5}},\ \bibinfo {pages} {332 EP } (\bibinfo {year} {2009})}\BibitemShut
  {NoStop}%
\bibitem [{\citenamefont {Pigeau}\ \emph {et~al.}(2010)\citenamefont {Pigeau},
  \citenamefont {de~Loubens}, \citenamefont {Klein}, \citenamefont {Riegler},
  \citenamefont {Lochner}, \citenamefont {Schmidt},\ and\ \citenamefont
  {Molenkamp}}]{Pigeau2010}%
  \BibitemOpen
  \bibfield  {author} {\bibinfo {author} {\bibfnamefont {B.}~\bibnamefont
  {Pigeau}}, \bibinfo {author} {\bibfnamefont {G.}~\bibnamefont {de~Loubens}},
  \bibinfo {author} {\bibfnamefont {O.}~\bibnamefont {Klein}}, \bibinfo
  {author} {\bibfnamefont {A.}~\bibnamefont {Riegler}}, \bibinfo {author}
  {\bibfnamefont {F.}~\bibnamefont {Lochner}}, \bibinfo {author} {\bibfnamefont
  {G.}~\bibnamefont {Schmidt}}, \ and\ \bibinfo {author} {\bibfnamefont
  {L.~W.}\ \bibnamefont {Molenkamp}},\ }\href
  {http://dx.doi.org/10.1038/nphys1810} {\bibfield  {journal} {\bibinfo
  {journal} {Nature Physics}\ }\textbf {\bibinfo {volume} {7}},\ \bibinfo
  {pages} {26 EP } (\bibinfo {year} {2010})}\BibitemShut {NoStop}%
\bibitem [{\citenamefont {Sugimoto}\ \emph {et~al.}(2011)\citenamefont
  {Sugimoto}, \citenamefont {Fukuma}, \citenamefont {Kasai}, \citenamefont
  {Kimura}, \citenamefont {Barman},\ and\ \citenamefont
  {Otani}}]{PhysRevLett.106.197203}%
  \BibitemOpen
  \bibfield  {author} {\bibinfo {author} {\bibfnamefont {S.}~\bibnamefont
  {Sugimoto}}, \bibinfo {author} {\bibfnamefont {Y.}~\bibnamefont {Fukuma}},
  \bibinfo {author} {\bibfnamefont {S.}~\bibnamefont {Kasai}}, \bibinfo
  {author} {\bibfnamefont {T.}~\bibnamefont {Kimura}}, \bibinfo {author}
  {\bibfnamefont {A.}~\bibnamefont {Barman}}, \ and\ \bibinfo {author}
  {\bibfnamefont {Y.}~\bibnamefont {Otani}},\ }\href {\doibase
  10.1103/PhysRevLett.106.197203} {\bibfield  {journal} {\bibinfo  {journal}
  {Phys. Rev. Lett.}\ }\textbf {\bibinfo {volume} {106}},\ \bibinfo {pages}
  {197203} (\bibinfo {year} {2011})}\BibitemShut {NoStop}%
\bibitem [{\citenamefont {Petit-Watelot}\ \emph {et~al.}(2012)\citenamefont
  {Petit-Watelot}, \citenamefont {Kim}, \citenamefont {Ruotolo}, \citenamefont
  {Otxoa}, \citenamefont {Bouzehouane}, \citenamefont {Grollier}, \citenamefont
  {Vansteenkiste}, \citenamefont {Van~de Wiele}, \citenamefont {Cros},\ and\
  \citenamefont {Devolder}}]{Petit-Watelot2012}%
  \BibitemOpen
  \bibfield  {author} {\bibinfo {author} {\bibfnamefont {S.}~\bibnamefont
  {Petit-Watelot}}, \bibinfo {author} {\bibfnamefont {J.-V.}\ \bibnamefont
  {Kim}}, \bibinfo {author} {\bibfnamefont {A.}~\bibnamefont {Ruotolo}},
  \bibinfo {author} {\bibfnamefont {R.~M.}\ \bibnamefont {Otxoa}}, \bibinfo
  {author} {\bibfnamefont {K.}~\bibnamefont {Bouzehouane}}, \bibinfo {author}
  {\bibfnamefont {J.}~\bibnamefont {Grollier}}, \bibinfo {author}
  {\bibfnamefont {A.}~\bibnamefont {Vansteenkiste}}, \bibinfo {author}
  {\bibfnamefont {B.}~\bibnamefont {Van~de Wiele}}, \bibinfo {author}
  {\bibfnamefont {V.}~\bibnamefont {Cros}}, \ and\ \bibinfo {author}
  {\bibfnamefont {T.}~\bibnamefont {Devolder}},\ }\href
  {http://dx.doi.org/10.1038/nphys2362} {\bibfield  {journal} {\bibinfo
  {journal} {Nature Physics}\ }\textbf {\bibinfo {volume} {8}},\ \bibinfo
  {pages} {682 EP } (\bibinfo {year} {2012})}\BibitemShut {NoStop}%
\bibitem [{\citenamefont {Sukhostavets}\ \emph {et~al.}(2013)\citenamefont
  {Sukhostavets}, \citenamefont {Pigeau}, \citenamefont {Sangiao},
  \citenamefont {de~Loubens}, \citenamefont {Naletov}, \citenamefont {Klein},
  \citenamefont {Mitsuzuka}, \citenamefont {Andrieu}, \citenamefont
  {Montaigne},\ and\ \citenamefont {Guslienko}}]{PhysRevLett.111.247601}%
  \BibitemOpen
  \bibfield  {author} {\bibinfo {author} {\bibfnamefont {O.~V.}\ \bibnamefont
  {Sukhostavets}}, \bibinfo {author} {\bibfnamefont {B.}~\bibnamefont
  {Pigeau}}, \bibinfo {author} {\bibfnamefont {S.}~\bibnamefont {Sangiao}},
  \bibinfo {author} {\bibfnamefont {G.}~\bibnamefont {de~Loubens}}, \bibinfo
  {author} {\bibfnamefont {V.~V.}\ \bibnamefont {Naletov}}, \bibinfo {author}
  {\bibfnamefont {O.}~\bibnamefont {Klein}}, \bibinfo {author} {\bibfnamefont
  {K.}~\bibnamefont {Mitsuzuka}}, \bibinfo {author} {\bibfnamefont
  {S.}~\bibnamefont {Andrieu}}, \bibinfo {author} {\bibfnamefont
  {F.}~\bibnamefont {Montaigne}}, \ and\ \bibinfo {author} {\bibfnamefont
  {K.~Y.}\ \bibnamefont {Guslienko}},\ }\href {\doibase
  10.1103/PhysRevLett.111.247601} {\bibfield  {journal} {\bibinfo  {journal}
  {Phys. Rev. Lett.}\ }\textbf {\bibinfo {volume} {111}},\ \bibinfo {pages}
  {247601} (\bibinfo {year} {2013})}\BibitemShut {NoStop}%
\bibitem [{\citenamefont {Noske}\ \emph {et~al.}(2016)\citenamefont {Noske},
  \citenamefont {Stoll}, \citenamefont {F\"ahnle}, \citenamefont {Gangwar},
  \citenamefont {Woltersdorf}, \citenamefont {Slavin}, \citenamefont {Weigand},
  \citenamefont {Dieterle}, \citenamefont {F\"orster}, \citenamefont {Back},\
  and\ \citenamefont {Sch\"utz}}]{PhysRevLett.117.037208}%
  \BibitemOpen
  \bibfield  {author} {\bibinfo {author} {\bibfnamefont {M.}~\bibnamefont
  {Noske}}, \bibinfo {author} {\bibfnamefont {H.}~\bibnamefont {Stoll}},
  \bibinfo {author} {\bibfnamefont {M.}~\bibnamefont {F\"ahnle}}, \bibinfo
  {author} {\bibfnamefont {A.}~\bibnamefont {Gangwar}}, \bibinfo {author}
  {\bibfnamefont {G.}~\bibnamefont {Woltersdorf}}, \bibinfo {author}
  {\bibfnamefont {A.}~\bibnamefont {Slavin}}, \bibinfo {author} {\bibfnamefont
  {M.}~\bibnamefont {Weigand}}, \bibinfo {author} {\bibfnamefont
  {G.}~\bibnamefont {Dieterle}}, \bibinfo {author} {\bibfnamefont
  {J.}~\bibnamefont {F\"orster}}, \bibinfo {author} {\bibfnamefont {C.~H.}\
  \bibnamefont {Back}}, \ and\ \bibinfo {author} {\bibfnamefont
  {G.}~\bibnamefont {Sch\"utz}},\ }\href {\doibase
  10.1103/PhysRevLett.117.037208} {\bibfield  {journal} {\bibinfo  {journal}
  {Phys. Rev. Lett.}\ }\textbf {\bibinfo {volume} {117}},\ \bibinfo {pages}
  {037208} (\bibinfo {year} {2016})}\BibitemShut {NoStop}%
\bibitem [{\citenamefont {Yamada}\ \emph {et~al.}(2007)\citenamefont {Yamada},
  \citenamefont {Kasai}, \citenamefont {Nakatani}, \citenamefont {Kobayashi},
  \citenamefont {Kohno}, \citenamefont {Thiaville},\ and\ \citenamefont
  {Ono}}]{Yamada2007}%
  \BibitemOpen
  \bibfield  {author} {\bibinfo {author} {\bibfnamefont {K.}~\bibnamefont
  {Yamada}}, \bibinfo {author} {\bibfnamefont {S.}~\bibnamefont {Kasai}},
  \bibinfo {author} {\bibfnamefont {Y.}~\bibnamefont {Nakatani}}, \bibinfo
  {author} {\bibfnamefont {K.}~\bibnamefont {Kobayashi}}, \bibinfo {author}
  {\bibfnamefont {H.}~\bibnamefont {Kohno}}, \bibinfo {author} {\bibfnamefont
  {A.}~\bibnamefont {Thiaville}}, \ and\ \bibinfo {author} {\bibfnamefont
  {T.}~\bibnamefont {Ono}},\ }\href {http://dx.doi.org/10.1038/nmat1867}
  {\bibfield  {journal} {\bibinfo  {journal} {Nature Materials}\ }\textbf
  {\bibinfo {volume} {6}},\ \bibinfo {pages} {270 EP } (\bibinfo {year}
  {2007})}\BibitemShut {NoStop}%
\bibitem [{\citenamefont {Bohlens}\ \emph {et~al.}(2008)\citenamefont
  {Bohlens}, \citenamefont {Kr{\"u}ger}, \citenamefont {Drews}, \citenamefont
  {Bolte}, \citenamefont {Meier},\ and\ \citenamefont
  {Pfannkuche}}]{doi:10.1063/1.2998584}%
  \BibitemOpen
  \bibfield  {author} {\bibinfo {author} {\bibfnamefont {S.}~\bibnamefont
  {Bohlens}}, \bibinfo {author} {\bibfnamefont {B.}~\bibnamefont {Kr{\"u}ger}},
  \bibinfo {author} {\bibfnamefont {A.}~\bibnamefont {Drews}}, \bibinfo
  {author} {\bibfnamefont {M.}~\bibnamefont {Bolte}}, \bibinfo {author}
  {\bibfnamefont {G.}~\bibnamefont {Meier}}, \ and\ \bibinfo {author}
  {\bibfnamefont {D.}~\bibnamefont {Pfannkuche}},\ }\href {\doibase
  10.1063/1.2998584} {\bibfield  {journal} {\bibinfo  {journal} {Applied
  Physics Letters}\ }\textbf {\bibinfo {volume} {93}},\ \bibinfo {pages}
  {142508} (\bibinfo {year} {2008})},\ \Eprint
  {http://arxiv.org/abs/http://dx.doi.org/10.1063/1.2998584}
  {http://dx.doi.org/10.1063/1.2998584} \BibitemShut {NoStop}%
\bibitem [{\citenamefont {Nakano}\ \emph {et~al.}(2011)\citenamefont {Nakano},
  \citenamefont {Chiba}, \citenamefont {Ohshima}, \citenamefont {Kasai},
  \citenamefont {Sato}, \citenamefont {Nakatani}, \citenamefont {Sekiguchi},
  \citenamefont {Kobayashi},\ and\ \citenamefont {Ono}}]{Nakano2011}%
  \BibitemOpen
  \bibfield  {author} {\bibinfo {author} {\bibfnamefont {K.}~\bibnamefont
  {Nakano}}, \bibinfo {author} {\bibfnamefont {D.}~\bibnamefont {Chiba}},
  \bibinfo {author} {\bibfnamefont {N.}~\bibnamefont {Ohshima}}, \bibinfo
  {author} {\bibfnamefont {S.}~\bibnamefont {Kasai}}, \bibinfo {author}
  {\bibfnamefont {T.}~\bibnamefont {Sato}}, \bibinfo {author} {\bibfnamefont
  {Y.}~\bibnamefont {Nakatani}}, \bibinfo {author} {\bibfnamefont
  {K.}~\bibnamefont {Sekiguchi}}, \bibinfo {author} {\bibfnamefont
  {K.}~\bibnamefont {Kobayashi}}, \ and\ \bibinfo {author} {\bibfnamefont
  {T.}~\bibnamefont {Ono}},\ }\href {\doibase 10.1063/1.3673303} {\bibfield
  {journal} {\bibinfo  {journal} {Applied Physics Letters}\ }\textbf {\bibinfo
  {volume} {99}},\ \bibinfo {pages} {262505} (\bibinfo {year} {2011})},\
  \Eprint {http://arxiv.org/abs/http://dx.doi.org/10.1063/1.3673303}
  {http://dx.doi.org/10.1063/1.3673303} \BibitemShut {NoStop}%
\bibitem [{\citenamefont {Yu}\ \emph {et~al.}(2011)\citenamefont {Yu},
  \citenamefont {Jung}, \citenamefont {Lee}, \citenamefont {Fischer},\ and\
  \citenamefont {Kim}}]{doi:10.1063/1.3551524}%
  \BibitemOpen
  \bibfield  {author} {\bibinfo {author} {\bibfnamefont {Y.-S.}\ \bibnamefont
  {Yu}}, \bibinfo {author} {\bibfnamefont {H.}~\bibnamefont {Jung}}, \bibinfo
  {author} {\bibfnamefont {K.-S.}\ \bibnamefont {Lee}}, \bibinfo {author}
  {\bibfnamefont {P.}~\bibnamefont {Fischer}}, \ and\ \bibinfo {author}
  {\bibfnamefont {S.-K.}\ \bibnamefont {Kim}},\ }\href {\doibase
  10.1063/1.3551524} {\bibfield  {journal} {\bibinfo  {journal} {Applied
  Physics Letters}\ }\textbf {\bibinfo {volume} {98}},\ \bibinfo {pages}
  {052507} (\bibinfo {year} {2011})},\ \Eprint
  {http://arxiv.org/abs/http://dx.doi.org/10.1063/1.3551524}
  {http://dx.doi.org/10.1063/1.3551524} \BibitemShut {NoStop}%
\bibitem [{\citenamefont {Geng}\ and\ \citenamefont {Jin}(2017)}]{Geng201784}%
  \BibitemOpen
  \bibfield  {author} {\bibinfo {author} {\bibfnamefont {L.~D.}\ \bibnamefont
  {Geng}}\ and\ \bibinfo {author} {\bibfnamefont {Y.~M.}\ \bibnamefont {Jin}},\
  }\href {\doibase 10.1016/j.jmmm.2016.09.062} {\bibfield  {journal} {\bibinfo
  {journal} {Journal of Magnetism and Magnetic Materials}\ }\textbf {\bibinfo
  {volume} {423}},\ \bibinfo {pages} {84} (\bibinfo {year} {2017})}\BibitemShut
  {NoStop}%
\bibitem [{\citenamefont {Velten}\ \emph {et~al.}(2017)\citenamefont {Velten},
  \citenamefont {Streubel}, \citenamefont {Farhan}, \citenamefont {Kent},
  \citenamefont {Im}, \citenamefont {Scholl}, \citenamefont {Dhuey},
  \citenamefont {Behncke}, \citenamefont {Meier},\ and\ \citenamefont
  {Fischer}}]{doi:10.1063/1.4990990}%
  \BibitemOpen
  \bibfield  {author} {\bibinfo {author} {\bibfnamefont {S.}~\bibnamefont
  {Velten}}, \bibinfo {author} {\bibfnamefont {R.}~\bibnamefont {Streubel}},
  \bibinfo {author} {\bibfnamefont {A.}~\bibnamefont {Farhan}}, \bibinfo
  {author} {\bibfnamefont {N.}~\bibnamefont {Kent}}, \bibinfo {author}
  {\bibfnamefont {M.-Y.}\ \bibnamefont {Im}}, \bibinfo {author} {\bibfnamefont
  {A.}~\bibnamefont {Scholl}}, \bibinfo {author} {\bibfnamefont
  {S.}~\bibnamefont {Dhuey}}, \bibinfo {author} {\bibfnamefont
  {C.}~\bibnamefont {Behncke}}, \bibinfo {author} {\bibfnamefont
  {G.}~\bibnamefont {Meier}}, \ and\ \bibinfo {author} {\bibfnamefont
  {P.}~\bibnamefont {Fischer}},\ }\href {\doibase 10.1063/1.4990990} {\bibfield
   {journal} {\bibinfo  {journal} {Applied Physics Letters}\ }\textbf {\bibinfo
  {volume} {110}},\ \bibinfo {pages} {262406} (\bibinfo {year} {2017})},\
  \Eprint {http://arxiv.org/abs/http://dx.doi.org/10.1063/1.4990990}
  {http://dx.doi.org/10.1063/1.4990990} \BibitemShut {NoStop}%
\bibitem [{\citenamefont {Shibata}\ and\ \citenamefont
  {Otani}(2004)}]{PhysRevB.70.012404}%
  \BibitemOpen
  \bibfield  {author} {\bibinfo {author} {\bibfnamefont {J.}~\bibnamefont
  {Shibata}}\ and\ \bibinfo {author} {\bibfnamefont {Y.}~\bibnamefont
  {Otani}},\ }\href {\doibase 10.1103/PhysRevB.70.012404} {\bibfield  {journal}
  {\bibinfo  {journal} {Phys. Rev. B}\ }\textbf {\bibinfo {volume} {70}},\
  \bibinfo {pages} {012404} (\bibinfo {year} {2004})}\BibitemShut {NoStop}%
\bibitem [{\citenamefont {Galkin}\ \emph {et~al.}(2006)\citenamefont {Galkin},
  \citenamefont {Ivanov},\ and\ \citenamefont {Zaspel}}]{PhysRevB.74.144419}%
  \BibitemOpen
  \bibfield  {author} {\bibinfo {author} {\bibfnamefont {A.~Y.}\ \bibnamefont
  {Galkin}}, \bibinfo {author} {\bibfnamefont {B.~A.}\ \bibnamefont {Ivanov}},
  \ and\ \bibinfo {author} {\bibfnamefont {C.~E.}\ \bibnamefont {Zaspel}},\
  }\href {\doibase 10.1103/PhysRevB.74.144419} {\bibfield  {journal} {\bibinfo
  {journal} {Phys. Rev. B}\ }\textbf {\bibinfo {volume} {74}},\ \bibinfo
  {pages} {144419} (\bibinfo {year} {2006})}\BibitemShut {NoStop}%
\bibitem [{\citenamefont {Pribiag}\ \emph {et~al.}(2007)\citenamefont
  {Pribiag}, \citenamefont {Krivorotov}, \citenamefont {Fuchs}, \citenamefont
  {Braganca}, \citenamefont {Ozatay}, \citenamefont {Sankey}, \citenamefont
  {Ralph},\ and\ \citenamefont {Buhrman}}]{Pribiag2007}%
  \BibitemOpen
  \bibfield  {author} {\bibinfo {author} {\bibfnamefont {V.~S.}\ \bibnamefont
  {Pribiag}}, \bibinfo {author} {\bibfnamefont {I.~N.}\ \bibnamefont
  {Krivorotov}}, \bibinfo {author} {\bibfnamefont {G.~D.}\ \bibnamefont
  {Fuchs}}, \bibinfo {author} {\bibfnamefont {P.~M.}\ \bibnamefont {Braganca}},
  \bibinfo {author} {\bibfnamefont {O.}~\bibnamefont {Ozatay}}, \bibinfo
  {author} {\bibfnamefont {J.~C.}\ \bibnamefont {Sankey}}, \bibinfo {author}
  {\bibfnamefont {D.~C.}\ \bibnamefont {Ralph}}, \ and\ \bibinfo {author}
  {\bibfnamefont {R.~A.}\ \bibnamefont {Buhrman}},\ }\href {\doibase
  10.1038/nphys619} {\bibfield  {journal} {\bibinfo  {journal} {Nat Phys}\
  }\textbf {\bibinfo {volume} {3}},\ \bibinfo {pages} {498} (\bibinfo {year}
  {2007})}\BibitemShut {NoStop}%
\bibitem [{\citenamefont {Mistral}\ \emph {et~al.}(2008)\citenamefont
  {Mistral}, \citenamefont {van Kampen}, \citenamefont {Hrkac}, \citenamefont
  {Kim}, \citenamefont {Devolder}, \citenamefont {Crozat}, \citenamefont
  {Chappert}, \citenamefont {Lagae},\ and\ \citenamefont
  {Schrefl}}]{PhysRevLett.100.257201}%
  \BibitemOpen
  \bibfield  {author} {\bibinfo {author} {\bibfnamefont {Q.}~\bibnamefont
  {Mistral}}, \bibinfo {author} {\bibfnamefont {M.}~\bibnamefont {van Kampen}},
  \bibinfo {author} {\bibfnamefont {G.}~\bibnamefont {Hrkac}}, \bibinfo
  {author} {\bibfnamefont {J.-V.}\ \bibnamefont {Kim}}, \bibinfo {author}
  {\bibfnamefont {T.}~\bibnamefont {Devolder}}, \bibinfo {author}
  {\bibfnamefont {P.}~\bibnamefont {Crozat}}, \bibinfo {author} {\bibfnamefont
  {C.}~\bibnamefont {Chappert}}, \bibinfo {author} {\bibfnamefont
  {L.}~\bibnamefont {Lagae}}, \ and\ \bibinfo {author} {\bibfnamefont
  {T.}~\bibnamefont {Schrefl}},\ }\href {\doibase
  10.1103/PhysRevLett.100.257201} {\bibfield  {journal} {\bibinfo  {journal}
  {Phys. Rev. Lett.}\ }\textbf {\bibinfo {volume} {100}},\ \bibinfo {pages}
  {257201} (\bibinfo {year} {2008})}\BibitemShut {NoStop}%
\bibitem [{\citenamefont {Soucaille}\ \emph {et~al.}(2017)\citenamefont
  {Soucaille}, \citenamefont {Kim}, \citenamefont {Devolder}, \citenamefont
  {Petit-Watelot}, \citenamefont {Manfrini}, \citenamefont {Roy},\ and\
  \citenamefont {Lagae}}]{0022-3727-50-8-085002}%
  \BibitemOpen
  \bibfield  {author} {\bibinfo {author} {\bibfnamefont {R.}~\bibnamefont
  {Soucaille}}, \bibinfo {author} {\bibfnamefont {J.-V.}\ \bibnamefont {Kim}},
  \bibinfo {author} {\bibfnamefont {T.}~\bibnamefont {Devolder}}, \bibinfo
  {author} {\bibfnamefont {S.}~\bibnamefont {Petit-Watelot}}, \bibinfo {author}
  {\bibfnamefont {M.}~\bibnamefont {Manfrini}}, \bibinfo {author}
  {\bibfnamefont {W.~V.}\ \bibnamefont {Roy}}, \ and\ \bibinfo {author}
  {\bibfnamefont {L.}~\bibnamefont {Lagae}},\ }\href
  {http://stacks.iop.org/0022-3727/50/i=8/a=085002} {\bibfield  {journal}
  {\bibinfo  {journal} {Journal of Physics D: Applied Physics}\ }\textbf
  {\bibinfo {volume} {50}},\ \bibinfo {pages} {085002} (\bibinfo {year}
  {2017})}\BibitemShut {NoStop}%
\bibitem [{\citenamefont {Kim}\ \emph {et~al.}(2010)\citenamefont {Kim},
  \citenamefont {Rozhkova}, \citenamefont {Ulasov}, \citenamefont {Bader},
  \citenamefont {Rajh}, \citenamefont {Lesniak},\ and\ \citenamefont
  {Novosad}}]{Kim2010}%
  \BibitemOpen
  \bibfield  {author} {\bibinfo {author} {\bibfnamefont {D.-H.}\ \bibnamefont
  {Kim}}, \bibinfo {author} {\bibfnamefont {E.~A.}\ \bibnamefont {Rozhkova}},
  \bibinfo {author} {\bibfnamefont {I.~V.}\ \bibnamefont {Ulasov}}, \bibinfo
  {author} {\bibfnamefont {S.~D.}\ \bibnamefont {Bader}}, \bibinfo {author}
  {\bibfnamefont {T.}~\bibnamefont {Rajh}}, \bibinfo {author} {\bibfnamefont
  {M.~S.}\ \bibnamefont {Lesniak}}, \ and\ \bibinfo {author} {\bibfnamefont
  {V.}~\bibnamefont {Novosad}},\ }\href {\doibase 10.1038/nmat2591} {\bibfield
  {journal} {\bibinfo  {journal} {Nat Mater}\ }\textbf {\bibinfo {volume}
  {9}},\ \bibinfo {pages} {165} (\bibinfo {year} {2010})}\BibitemShut {NoStop}%
\bibitem [{\citenamefont {Wong}\ \emph {et~al.}(2017)\citenamefont {Wong},
  \citenamefont {Gan}, \citenamefont {Liu},\ and\ \citenamefont
  {Lew}}]{Wong2017}%
  \BibitemOpen
  \bibfield  {author} {\bibinfo {author} {\bibfnamefont {D.~W.}\ \bibnamefont
  {Wong}}, \bibinfo {author} {\bibfnamefont {W.~L.}\ \bibnamefont {Gan}},
  \bibinfo {author} {\bibfnamefont {N.}~\bibnamefont {Liu}}, \ and\ \bibinfo
  {author} {\bibfnamefont {W.~S.}\ \bibnamefont {Lew}},\ }\href {\doibase
  10.1038/s41598-017-11279-w} {\bibfield  {journal} {\bibinfo  {journal}
  {Scientific Reports}\ }\textbf {\bibinfo {volume} {7}},\ \bibinfo {pages}
  {10919} (\bibinfo {year} {2017})}\BibitemShut {NoStop}%
\bibitem [{\citenamefont {Cherepov}\ \emph {et~al.}(2012)\citenamefont
  {Cherepov}, \citenamefont {Koop}, \citenamefont {Galkin}, \citenamefont
  {Khymyn}, \citenamefont {Ivanov}, \citenamefont {Worledge},\ and\
  \citenamefont {Korenivski}}]{PhysRevLett.109.097204}%
  \BibitemOpen
  \bibfield  {author} {\bibinfo {author} {\bibfnamefont {S.~S.}\ \bibnamefont
  {Cherepov}}, \bibinfo {author} {\bibfnamefont {B.~C.}\ \bibnamefont {Koop}},
  \bibinfo {author} {\bibfnamefont {A.~Y.}\ \bibnamefont {Galkin}}, \bibinfo
  {author} {\bibfnamefont {R.~S.}\ \bibnamefont {Khymyn}}, \bibinfo {author}
  {\bibfnamefont {B.~A.}\ \bibnamefont {Ivanov}}, \bibinfo {author}
  {\bibfnamefont {D.~C.}\ \bibnamefont {Worledge}}, \ and\ \bibinfo {author}
  {\bibfnamefont {V.}~\bibnamefont {Korenivski}},\ }\href {\doibase
  10.1103/PhysRevLett.109.097204} {\bibfield  {journal} {\bibinfo  {journal}
  {Phys. Rev. Lett.}\ }\textbf {\bibinfo {volume} {109}},\ \bibinfo {pages}
  {097204} (\bibinfo {year} {2012})}\BibitemShut {NoStop}%
\bibitem [{\citenamefont {H{\"a}nze}\ \emph {et~al.}(2016)\citenamefont
  {H{\"a}nze}, \citenamefont {Adolff}, \citenamefont {Velten}, \citenamefont
  {Weigand},\ and\ \citenamefont {Meier}}]{PhysRevB.93.054411}%
  \BibitemOpen
  \bibfield  {author} {\bibinfo {author} {\bibfnamefont {M.}~\bibnamefont
  {H{\"a}nze}}, \bibinfo {author} {\bibfnamefont {C.~F.}\ \bibnamefont
  {Adolff}}, \bibinfo {author} {\bibfnamefont {S.}~\bibnamefont {Velten}},
  \bibinfo {author} {\bibfnamefont {M.}~\bibnamefont {Weigand}}, \ and\
  \bibinfo {author} {\bibfnamefont {G.}~\bibnamefont {Meier}},\ }\href
  {\doibase 10.1103/PhysRevB.93.054411} {\bibfield  {journal} {\bibinfo
  {journal} {Phys. Rev. B}\ }\textbf {\bibinfo {volume} {93}},\ \bibinfo
  {pages} {054411} (\bibinfo {year} {2016})}\BibitemShut {NoStop}%
\bibitem [{\citenamefont {Stebliy}\ \emph {et~al.}(2017)\citenamefont
  {Stebliy}, \citenamefont {Jain}, \citenamefont {Kolesnikov}, \citenamefont
  {Ognev}, \citenamefont {Samardak}, \citenamefont {Davydenko}, \citenamefont
  {Sukovatitcina}, \citenamefont {Chebotkevich}, \citenamefont {Ding},
  \citenamefont {Pearson}, \citenamefont {Khovaylo},\ and\ \citenamefont
  {Novosad}}]{Stebliy2017}%
  \BibitemOpen
  \bibfield  {author} {\bibinfo {author} {\bibfnamefont {M.~E.}\ \bibnamefont
  {Stebliy}}, \bibinfo {author} {\bibfnamefont {S.}~\bibnamefont {Jain}},
  \bibinfo {author} {\bibfnamefont {A.~G.}\ \bibnamefont {Kolesnikov}},
  \bibinfo {author} {\bibfnamefont {A.~V.}\ \bibnamefont {Ognev}}, \bibinfo
  {author} {\bibfnamefont {A.~S.}\ \bibnamefont {Samardak}}, \bibinfo {author}
  {\bibfnamefont {A.~V.}\ \bibnamefont {Davydenko}}, \bibinfo {author}
  {\bibfnamefont {E.~V.}\ \bibnamefont {Sukovatitcina}}, \bibinfo {author}
  {\bibfnamefont {L.~A.}\ \bibnamefont {Chebotkevich}}, \bibinfo {author}
  {\bibfnamefont {J.}~\bibnamefont {Ding}}, \bibinfo {author} {\bibfnamefont
  {J.}~\bibnamefont {Pearson}}, \bibinfo {author} {\bibfnamefont
  {V.}~\bibnamefont {Khovaylo}}, \ and\ \bibinfo {author} {\bibfnamefont
  {V.}~\bibnamefont {Novosad}},\ }\href {\doibase 10.1038/s41598-017-01222-4}
  {\bibfield  {journal} {\bibinfo  {journal} {Scientific Reports}\ }\textbf
  {\bibinfo {volume} {7}},\ \bibinfo {pages} {1127} (\bibinfo {year}
  {2017})}\BibitemShut {NoStop}%
\bibitem [{\citenamefont {Thiele}(1973)}]{PhysRevLett.30.230}%
  \BibitemOpen
  \bibfield  {author} {\bibinfo {author} {\bibfnamefont {A.~A.}\ \bibnamefont
  {Thiele}},\ }\href {\doibase 10.1103/PhysRevLett.30.230} {\bibfield
  {journal} {\bibinfo  {journal} {Phys. Rev. Lett.}\ }\textbf {\bibinfo
  {volume} {30}},\ \bibinfo {pages} {230} (\bibinfo {year} {1973})}\BibitemShut
  {NoStop}%
\bibitem [{\citenamefont {Warden}\ and\ \citenamefont
  {Waldner}(1988)}]{doi:10.1063/1.342378}%
  \BibitemOpen
  \bibfield  {author} {\bibinfo {author} {\bibfnamefont {M.}~\bibnamefont
  {Warden}}\ and\ \bibinfo {author} {\bibfnamefont {F.}~\bibnamefont
  {Waldner}},\ }\href {\doibase 10.1063/1.342378} {\bibfield  {journal}
  {\bibinfo  {journal} {Journal of Applied Physics}\ }\textbf {\bibinfo
  {volume} {64}},\ \bibinfo {pages} {5386} (\bibinfo {year} {1988})},\ \Eprint
  {http://arxiv.org/abs/http://dx.doi.org/10.1063/1.342378}
  {http://dx.doi.org/10.1063/1.342378} \BibitemShut {NoStop}%
\bibitem [{\citenamefont {Wigen}\ \emph {et~al.}(1988)\citenamefont {Wigen},
  \citenamefont {Doetsch}, \citenamefont {Ming}, \citenamefont {Baselgia},\
  and\ \citenamefont {Waldner}}]{doi:10.1063/1.340525}%
  \BibitemOpen
  \bibfield  {author} {\bibinfo {author} {\bibfnamefont {P.~E.}\ \bibnamefont
  {Wigen}}, \bibinfo {author} {\bibfnamefont {H.}~\bibnamefont {Doetsch}},
  \bibinfo {author} {\bibfnamefont {Y.}~\bibnamefont {Ming}}, \bibinfo {author}
  {\bibfnamefont {L.}~\bibnamefont {Baselgia}}, \ and\ \bibinfo {author}
  {\bibfnamefont {F.}~\bibnamefont {Waldner}},\ }\href {\doibase
  10.1063/1.340525} {\bibfield  {journal} {\bibinfo  {journal} {Journal of
  Applied Physics}\ }\textbf {\bibinfo {volume} {63}},\ \bibinfo {pages} {4157}
  (\bibinfo {year} {1988})},\ \Eprint
  {http://arxiv.org/abs/http://dx.doi.org/10.1063/1.340525}
  {http://dx.doi.org/10.1063/1.340525} \BibitemShut {NoStop}%
\bibitem [{\citenamefont {{\'A}lvarez}\ \emph {et~al.}(2000)\citenamefont
  {{\'A}lvarez}, \citenamefont {Pla},\ and\ \citenamefont
  {Chubykalo}}]{PhysRevB.61.11613}%
  \BibitemOpen
  \bibfield  {author} {\bibinfo {author} {\bibfnamefont {L.~F.}\ \bibnamefont
  {{\'A}lvarez}}, \bibinfo {author} {\bibfnamefont {O.}~\bibnamefont {Pla}}, \
  and\ \bibinfo {author} {\bibfnamefont {O.}~\bibnamefont {Chubykalo}},\ }\href
  {\doibase 10.1103/PhysRevB.61.11613} {\bibfield  {journal} {\bibinfo
  {journal} {Phys. Rev. B}\ }\textbf {\bibinfo {volume} {61}},\ \bibinfo
  {pages} {11613} (\bibinfo {year} {2000})}\BibitemShut {NoStop}%
\bibitem [{\citenamefont {Dykman}\ \emph {et~al.}(2001)\citenamefont {Dykman},
  \citenamefont {Golding}, \citenamefont {McCann}, \citenamefont {Smelyanskiy},
  \citenamefont {Luchinsky}, \citenamefont {Mannella},\ and\ \citenamefont
  {McClintock}}]{doi:10.1063/1.1380368}%
  \BibitemOpen
  \bibfield  {author} {\bibinfo {author} {\bibfnamefont {M.~I.}\ \bibnamefont
  {Dykman}}, \bibinfo {author} {\bibfnamefont {B.}~\bibnamefont {Golding}},
  \bibinfo {author} {\bibfnamefont {L.~I.}\ \bibnamefont {McCann}}, \bibinfo
  {author} {\bibfnamefont {V.~N.}\ \bibnamefont {Smelyanskiy}}, \bibinfo
  {author} {\bibfnamefont {D.~G.}\ \bibnamefont {Luchinsky}}, \bibinfo {author}
  {\bibfnamefont {R.}~\bibnamefont {Mannella}}, \ and\ \bibinfo {author}
  {\bibfnamefont {P.~V.~E.}\ \bibnamefont {McClintock}},\ }\href {\doibase
  10.1063/1.1380368} {\bibfield  {journal} {\bibinfo  {journal} {Chaos: An
  Interdisciplinary Journal of Nonlinear Science}\ }\textbf {\bibinfo {volume}
  {11}},\ \bibinfo {pages} {587} (\bibinfo {year} {2001})},\ \Eprint
  {http://arxiv.org/abs/http://dx.doi.org/10.1063/1.1380368}
  {http://dx.doi.org/10.1063/1.1380368} \BibitemShut {NoStop}%
\bibitem [{\citenamefont {Devoret}\ \emph {et~al.}(1987)\citenamefont
  {Devoret}, \citenamefont {Esteve}, \citenamefont {Martinis}, \citenamefont
  {Cleland},\ and\ \citenamefont {Clarke}}]{Devoret1987}%
  \BibitemOpen
  \bibfield  {author} {\bibinfo {author} {\bibfnamefont {M.}~\bibnamefont
  {Devoret}}, \bibinfo {author} {\bibfnamefont {D.}~\bibnamefont {Esteve}},
  \bibinfo {author} {\bibfnamefont {J.}~\bibnamefont {Martinis}}, \bibinfo
  {author} {\bibfnamefont {A.}~\bibnamefont {Cleland}}, \ and\ \bibinfo
  {author} {\bibfnamefont {J.}~\bibnamefont {Clarke}},\ }\href
  {https://journals.aps.org/prb/abstract/10.1103/PhysRevB.36.58} {\bibfield
  {journal} {\bibinfo  {journal} {Physical Review B}\ }\textbf {\bibinfo
  {volume} {36}} (\bibinfo {year} {1987})}\BibitemShut {NoStop}%
\bibitem [{\citenamefont {Crisanti}\ \emph {et~al.}(1994)\citenamefont
  {Crisanti}, \citenamefont {Falcioni}, \citenamefont {Paladin},\ and\
  \citenamefont {Vulpiani}}]{0305-4470-27-17-001}%
  \BibitemOpen
  \bibfield  {author} {\bibinfo {author} {\bibfnamefont {A.}~\bibnamefont
  {Crisanti}}, \bibinfo {author} {\bibfnamefont {M.}~\bibnamefont {Falcioni}},
  \bibinfo {author} {\bibfnamefont {G.}~\bibnamefont {Paladin}}, \ and\
  \bibinfo {author} {\bibfnamefont {A.}~\bibnamefont {Vulpiani}},\ }\href
  {http://stacks.iop.org/0305-4470/27/i=17/a=001} {\bibfield  {journal}
  {\bibinfo  {journal} {Journal of Physics A: Mathematical and General}\
  }\textbf {\bibinfo {volume} {27}},\ \bibinfo {pages} {L597} (\bibinfo {year}
  {1994})}\BibitemShut {NoStop}%
\bibitem [{\citenamefont {Gammaitoni}\ \emph {et~al.}(1998)\citenamefont
  {Gammaitoni}, \citenamefont {H{\"a}nggi}, \citenamefont {Jung},\ and\
  \citenamefont {Marchesoni}}]{RevModPhys.70.223}%
  \BibitemOpen
  \bibfield  {author} {\bibinfo {author} {\bibfnamefont {L.}~\bibnamefont
  {Gammaitoni}}, \bibinfo {author} {\bibfnamefont {P.}~\bibnamefont
  {H{\"a}nggi}}, \bibinfo {author} {\bibfnamefont {P.}~\bibnamefont {Jung}}, \
  and\ \bibinfo {author} {\bibfnamefont {F.}~\bibnamefont {Marchesoni}},\
  }\href {\doibase 10.1103/RevModPhys.70.223} {\bibfield  {journal} {\bibinfo
  {journal} {Rev. Mod. Phys.}\ }\textbf {\bibinfo {volume} {70}},\ \bibinfo
  {pages} {223} (\bibinfo {year} {1998})}\BibitemShut {NoStop}%
\bibitem [{\citenamefont {Iansiti}\ \emph {et~al.}(1985)\citenamefont
  {Iansiti}, \citenamefont {Hu}, \citenamefont {Westervelt},\ and\
  \citenamefont {Tinkham}}]{PhysRevLett.55.746}%
  \BibitemOpen
  \bibfield  {author} {\bibinfo {author} {\bibfnamefont {M.}~\bibnamefont
  {Iansiti}}, \bibinfo {author} {\bibfnamefont {Q.}~\bibnamefont {Hu}},
  \bibinfo {author} {\bibfnamefont {R.~M.}\ \bibnamefont {Westervelt}}, \ and\
  \bibinfo {author} {\bibfnamefont {M.}~\bibnamefont {Tinkham}},\ }\href
  {\doibase 10.1103/PhysRevLett.55.746} {\bibfield  {journal} {\bibinfo
  {journal} {Phys. Rev. Lett.}\ }\textbf {\bibinfo {volume} {55}},\ \bibinfo
  {pages} {746} (\bibinfo {year} {1985})}\BibitemShut {NoStop}%
\bibitem [{\citenamefont {Kautz}(1985)}]{doi:10.1063/1.335642}%
  \BibitemOpen
  \bibfield  {author} {\bibinfo {author} {\bibfnamefont {R.~L.}\ \bibnamefont
  {Kautz}},\ }\href {\doibase 10.1063/1.335642} {\bibfield  {journal} {\bibinfo
   {journal} {Journal of Applied Physics}\ }\textbf {\bibinfo {volume} {58}},\
  \bibinfo {pages} {424} (\bibinfo {year} {1985})},\ \Eprint
  {http://arxiv.org/abs/https://doi.org/10.1063/1.335642}
  {https://doi.org/10.1063/1.335642} \BibitemShut {NoStop}%
\bibitem [{\citenamefont {Blackburn}\ \emph {et~al.}(1996)\citenamefont
  {Blackburn}, \citenamefont {Smith},\ and\ \citenamefont
  {Gr\o{}nbech-Jensen}}]{PhysRevB.53.14546}%
  \BibitemOpen
  \bibfield  {author} {\bibinfo {author} {\bibfnamefont {J.~A.}\ \bibnamefont
  {Blackburn}}, \bibinfo {author} {\bibfnamefont {H.~J.~T.}\ \bibnamefont
  {Smith}}, \ and\ \bibinfo {author} {\bibfnamefont {N.}~\bibnamefont
  {Gr\o{}nbech-Jensen}},\ }\href {\doibase 10.1103/PhysRevB.53.14546}
  {\bibfield  {journal} {\bibinfo  {journal} {Phys. Rev. B}\ }\textbf {\bibinfo
  {volume} {53}},\ \bibinfo {pages} {14546} (\bibinfo {year}
  {1996})}\BibitemShut {NoStop}%
\bibitem [{\citenamefont {Gaidis}\ \emph {et~al.}(2006)\citenamefont {Gaidis},
  \citenamefont {O'Sullivan}, \citenamefont {Nowak}, \citenamefont {Lu},
  \citenamefont {Kanakasabapathy}, \citenamefont {Trouilloud}, \citenamefont
  {Worledge}, \citenamefont {Assefa}, \citenamefont {Milkove}, \citenamefont
  {Wright},\ and\ \citenamefont {Gallagher}}]{Gaidis2006}%
  \BibitemOpen
  \bibfield  {author} {\bibinfo {author} {\bibfnamefont {M.~C.}\ \bibnamefont
  {Gaidis}}, \bibinfo {author} {\bibfnamefont {E.~J.}\ \bibnamefont
  {O'Sullivan}}, \bibinfo {author} {\bibfnamefont {J.~J.}\ \bibnamefont
  {Nowak}}, \bibinfo {author} {\bibfnamefont {Y.}~\bibnamefont {Lu}}, \bibinfo
  {author} {\bibfnamefont {S.}~\bibnamefont {Kanakasabapathy}}, \bibinfo
  {author} {\bibfnamefont {P.~L.}\ \bibnamefont {Trouilloud}}, \bibinfo
  {author} {\bibfnamefont {D.~C.}\ \bibnamefont {Worledge}}, \bibinfo {author}
  {\bibfnamefont {S.}~\bibnamefont {Assefa}}, \bibinfo {author} {\bibfnamefont
  {K.~R.}\ \bibnamefont {Milkove}}, \bibinfo {author} {\bibfnamefont {G.~P.}\
  \bibnamefont {Wright}}, \ and\ \bibinfo {author} {\bibfnamefont {W.~J.}\
  \bibnamefont {Gallagher}},\ }\href {\doibase 10.1147/rd.501.0041} {\bibfield
  {journal} {\bibinfo  {journal} {IBM Journal of Research and Development}\
  }\textbf {\bibinfo {volume} {50}},\ \bibinfo {pages} {41} (\bibinfo {year}
  {2006})}\BibitemShut {NoStop}%
\bibitem [{\citenamefont {Konovalenko}\ \emph {et~al.}(2009)\citenamefont
  {Konovalenko}, \citenamefont {Lindgren}, \citenamefont {Cherepov},
  \citenamefont {Korenivski},\ and\ \citenamefont
  {Worledge}}]{PhysRevB.80.144425}%
  \BibitemOpen
  \bibfield  {author} {\bibinfo {author} {\bibfnamefont {A.}~\bibnamefont
  {Konovalenko}}, \bibinfo {author} {\bibfnamefont {E.}~\bibnamefont
  {Lindgren}}, \bibinfo {author} {\bibfnamefont {S.~S.}\ \bibnamefont
  {Cherepov}}, \bibinfo {author} {\bibfnamefont {V.}~\bibnamefont
  {Korenivski}}, \ and\ \bibinfo {author} {\bibfnamefont {D.~C.}\ \bibnamefont
  {Worledge}},\ }\href {\doibase 10.1103/PhysRevB.80.144425} {\bibfield
  {journal} {\bibinfo  {journal} {Phys. Rev. B}\ }\textbf {\bibinfo {volume}
  {80}},\ \bibinfo {pages} {144425} (\bibinfo {year} {2009})}\BibitemShut
  {NoStop}%
\bibitem [{\citenamefont {Guslienko}(2006)}]{doi:10.1063/1.2221904}%
  \BibitemOpen
  \bibfield  {author} {\bibinfo {author} {\bibfnamefont {K.~Y.}\ \bibnamefont
  {Guslienko}},\ }\href {\doibase 10.1063/1.2221904} {\bibfield  {journal}
  {\bibinfo  {journal} {Applied Physics Letters}\ }\textbf {\bibinfo {volume}
  {89}},\ \bibinfo {pages} {022510} (\bibinfo {year} {2006})},\ \Eprint
  {http://arxiv.org/abs/https://doi.org/10.1063/1.2221904}
  {https://doi.org/10.1063/1.2221904} \BibitemShut {NoStop}%
\bibitem [{\citenamefont {Magiera}(2013)}]{0295-5075-103-5-57004}%
  \BibitemOpen
  \bibfield  {author} {\bibinfo {author} {\bibfnamefont {M.~P.}\ \bibnamefont
  {Magiera}},\ }\href {http://stacks.iop.org/0295-5075/103/i=5/a=57004}
  {\bibfield  {journal} {\bibinfo  {journal} {EPL (Europhysics Letters)}\
  }\textbf {\bibinfo {volume} {103}},\ \bibinfo {pages} {57004} (\bibinfo
  {year} {2013})}\BibitemShut {NoStop}%
\end{thebibliography}%
 
\end{document}


\title{Dynamic chaos in spin-vortex pairs: Supplementary material}

\author{A. Bondarenko}
\affiliation{Royal Institute of Technology, 10691 Stockholm, Sweden}
\affiliation{Institute of Magnetism, National Academy of Science, 03142 Kiev, Ukraine}

\author{E. Holmgren}
\author{Z. W. Li}
\affiliation{Royal Institute of Technology, 10691 Stockholm, Sweden}

\author{B. A. Ivanov}
\affiliation{Institute of Magnetism, National Academy of Science, 03142 Kiev, Ukraine}
\affiliation{National University of Science and Technology «MISiS», Moscow, 119049, Russian Federation}

\author{V. Korenivski}
\affiliation{Royal Institute of Technology, 10691 Stockholm, Sweden}


\maketitle


\section{Samples and measurement setup}
The samples studied are nanopillars with the free synthetic antiferromagnetic (SAF) layer composed of two symmetric Permalloy (Py, Ni$_{80}$Fe$_{20}$) layers, each 5 nm thick, separated by a TaN spacer of thickness 1 nm. The spacer material was chosen to suppress the inter-Py direct and indirect exchange coupling, such that the two Py layers interact only magnetostatically. The read-out of the SAF magnetic state is via the magnetoresistance of a 1 nm thick Al-O$_x$ tunnel junction separating the free-SAF and a reference pinned-SAF (see inset to Fig.\ref{Fig2}). The magnetic tunnel junction resistance is 1-2k$\Omega$ and the magnetoresistance about 20\%, measured between the two uniformly magnetized ground states, with the bottom free-SAF layer parallel and antiparallel to the top reference-SAF layer. The reference SAF, an exchange-biased CoFeB/Ru/CoFeB trilayer with the Ru thickness chosen to maximize the antiparallel-RKKY coupling, is nearly ideally flux-closed so as to produce no dc field bias in the free-SAF layer. 

\begin{figure}[!t]
\includegraphics[width=3in]{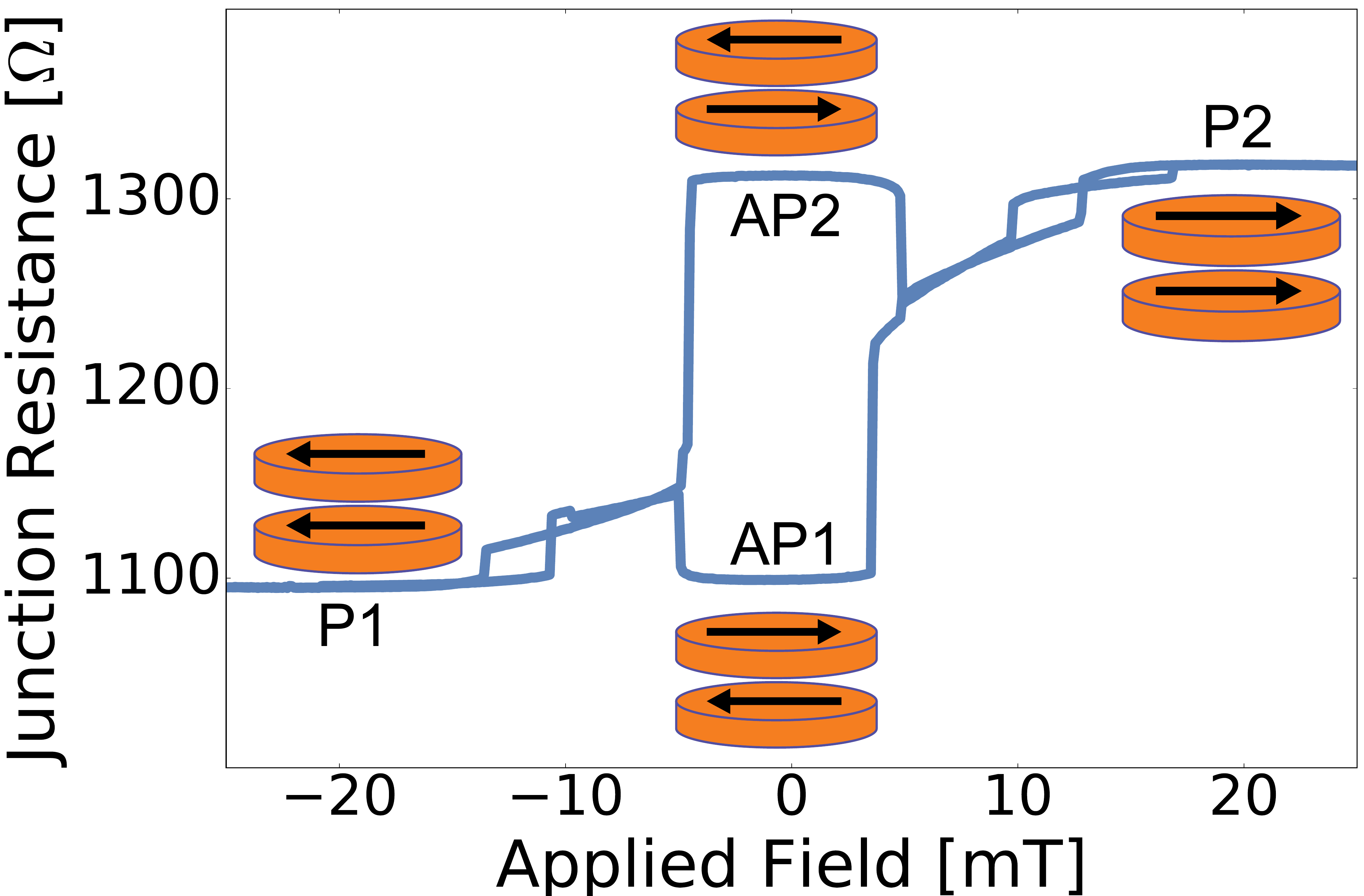}
\caption{Magnetoresistance of the uniform groundstate of a representative junction of size 350x420 nm, measured at 77 K. In the ground state the magnetization of the free layers are uniform and antiparallel (AP1,2 states), shown  schematically at different points in the field sweep with the in-plane arrows indicating the orientation of the individual Py particles.}
\label{Fig1}
\end{figure}

\begin{figure}[!t]
\includegraphics[width=3in]{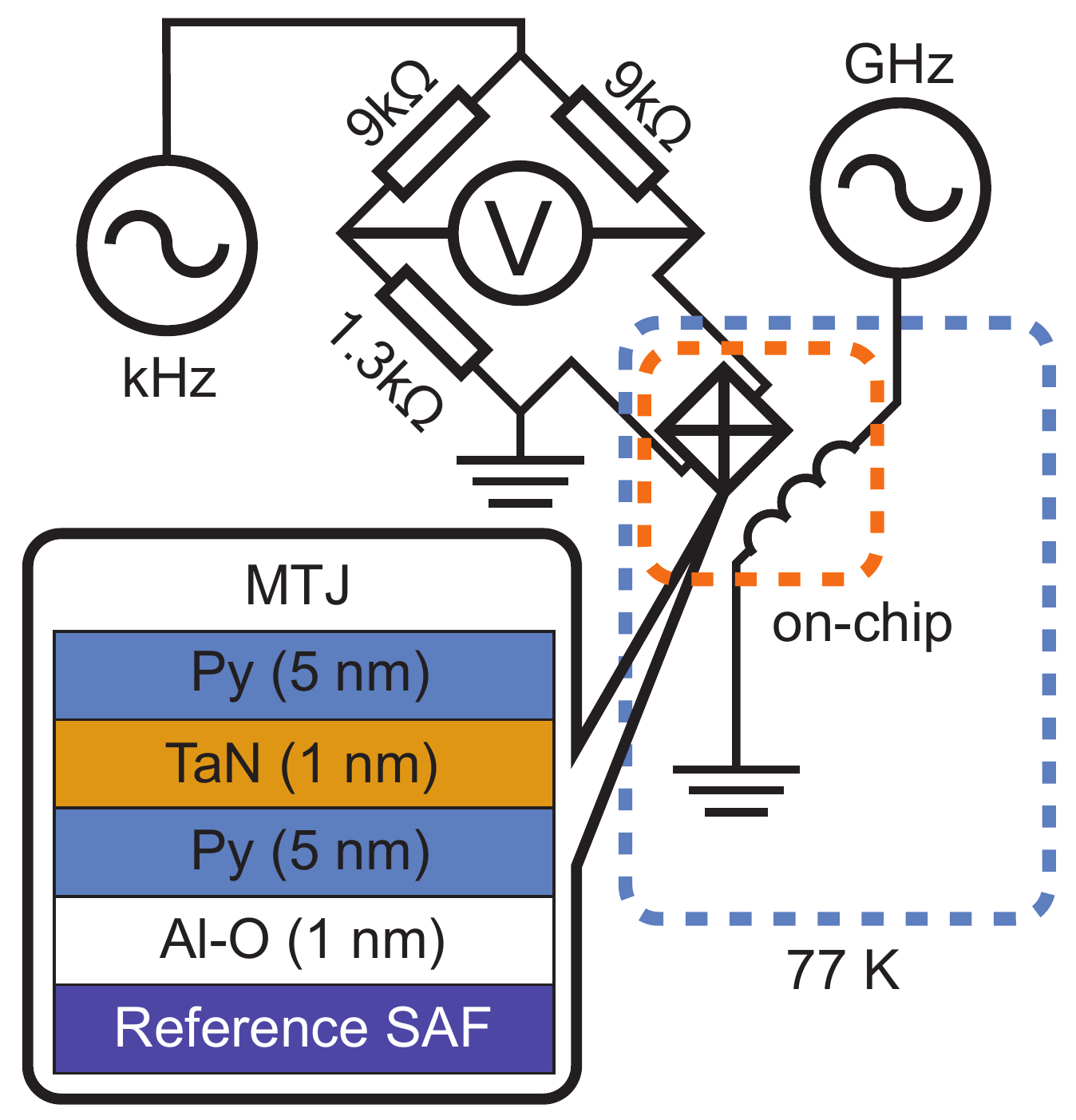}
\caption{Circuit diagram of the measurement setup. The junction resistance is readout using a Wheatstone-bridge, with the resistor values 9k$\Omega$ and 1.3k$\Omega$ in one arm and 9k$\Omega$ together with the junction in the other arm. GHz-range excitations are supplied to the junction through integrated 50$\Omega$ high-frequency lines connected to an rf-generator (Keysight 13 GHz) or an arbitrary-wave-generator (Tektronix 5 GS/s).}
\label{Fig2}
\end{figure}

The nanopillars were fabricated with the lateral dimensions ranging from 350x420 to 420x500 nm, designed to have the in-plane shape anisotropy of the individual Py layers of a few mT in the uniform-magnetization states of the junctions. A representative magnetoresistance loop is shown in Fig.\ref{Fig1}, illustrating the zero-field bi-stability of the free-SAF ground state (AP1 and AP2 states) and the typical spin-flip transitions at a few mT with subsequent saturation (P1 and P2 states) at about 20 mT. The  nanopillars are on-chip integrated in a toggle-style memory cell layout, as described in \footnote{W.J. Gallagher, S.S.P. Parkin, IBM J. Res. Dev. 50, 1 (2006).}, with the resistive readout electrically separated from the 50$\Omega$ high-frequency Cu line used to supply the GHz-range excitations, aligned such that the rf-field is at 45 degrees to the long axis of the SAF. 

The leads are terminated at the surface of the Si chip where they are wire-bonded to a chip-carrier and placed in a liquid-nitrogen bath cryostat fitted with a large external solenoid electro-magnet for suppling dc fields oriented along the long axis of the SAF. The junction resistance was measured using a Wheatstone bridge with the voltage measured by a Stanford Research Instruments SR830 DSP lock-in amplifier. The circuit diagram of the measurement setup is shown in Fig.\ref{Fig2}. The samples were fabricated using methods described in \footnote{M. C. Gaidis, et al., IBM J. Res. Dev. 50, 41 (2006).}.\\

\begin{figure*}[!t]
\includegraphics[width=6in]{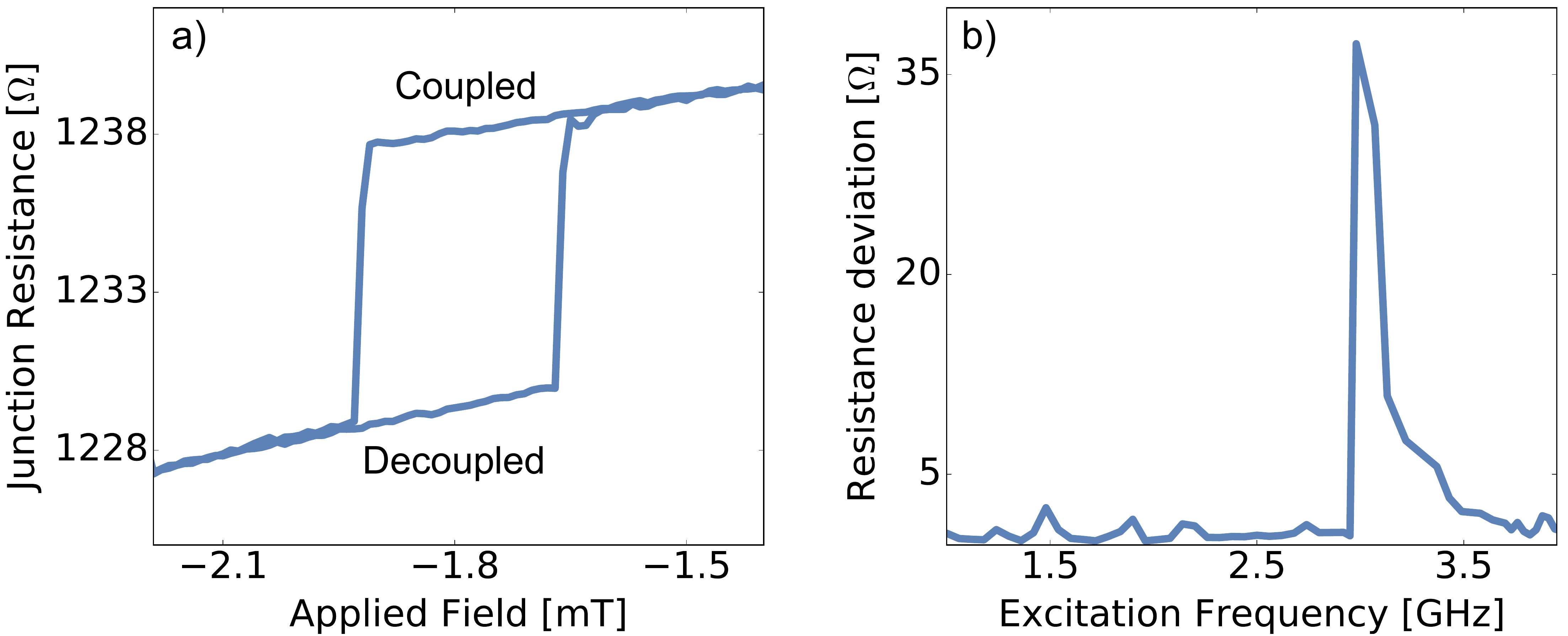}
\caption{(a) $R-H$ hysteresis loop corresponding to core coupling/decoupling. At small fields the cores of the P-AP vortex pair measured are centered in the Py particles and strongly bound. As the dc field increases the cores are pulled apart due to the antiparallel chirality. Above some threshold, roughly -2 mT, the cores decouple and act as individual cores. (b) Microwave spectrum of a P-AP vortex pair measured with the excitation amplitude of 0.3 mT (small-signal linear regime). The peak of the rotational resonance corresponds to an anti-phase rotation of the two cores about the pair's 'center-of-mass'. The frequency of the oscillation is sensitive to the lateral separation of the two cores and decreases as the separation is increased, which leads to a widening of the resonance peak toward lower frequencies as the rf field amplitude is increased (not shown).}
\label{Fig3}
\end{figure*}

\section{Measurement methods}
The vortex pair state was set in using the high-power high-frequency excitations with amplitudes of tens of mT and frequency in the range of 1-4 GHz. Once the vortex state is set, its lifetime at zero field is essentially infinite. In the vortex pair state 16 possible combinations of core-polarization and chirality can occur. Assuming the layers are symmetric these reduce to 4 non-degenerate states: Parallel core polarization and Parallel chirality (P-P), AP-AP, AP-P, and P-AP. The focus of this work is the non-linear dynamics of the P-AP state, signified by its strong core-core coupling combined with effective core-separation control. The type of a vortex pair produced was determine from the $R-H$ sweeps, in which the resistance change is related to the displacement of the bottom vortex core along the short axis of the junction. The key signature of a P-AP vortex pair is hysteresis in $R-H$, corresponding to coupling and decoupling of the cores. Further, the P-AP state is the only state with a resonance at about 3 GHz, with the two cores strongly coupled and rotating about the pair's center. A field sweep and microwave spectrum measured at 77 K for a typical P-AP spin vortex pair are shown in Fig.\ref{Fig3}. We have previously reported of the dynamic properties on the strongly-coupled P-AP state, in its quasi-linear regime, at room-temperature\footnote{S.S. Cherepov, et al., Phys. Rev. Lett. 109, 097204 (2012).}. Here we focus on the highly nonlinear, hysteretic core decoupling/recoupling dynamics, unmasked from thermal agitation by measurements at lower temperatures. \\ 

The measurements were performed by biasing the junction with a dc field to the center of the core-core hysteresis, -1.8 mT in Fig.\ref{Fig3}, then applying various rf excitations using the on-chip 50$\Omega$ lines and measuring the junction resistance to determine whether the system had switched between the two core-core states. If core-decoupling had occurred, the system was reset by toggling the static field. Two types of high-frequency excitations were used: continuous wave (CW) produced by a Keysight N5173B EXG rf generator and waveforms with precisely controlled number of periods produced by a Tektronix AWG 7052 arbitrary waveform generator. The lifetime of the coupled/decoupled states near the center of the hysteresis are infinite in relation to the measurement times used.

The CW signals had a typical duration of 300 ms (uncontrolled but roughly 10$^{11}$ periods), much longer than any intrinsic relaxation time in the measured system, ensuring a steady-state regime. The CW measurements were used for recording the amplitude-frequency maps, such as in Fig.2, with repetitions to ensure proper statistics. Prior to measurements, the rf feed lines, including the on-chip wire-bonding, were characterized and compensated for reflections via the rf generator sequence.

\begin{figure*}[!t]
\includegraphics[width=7in]{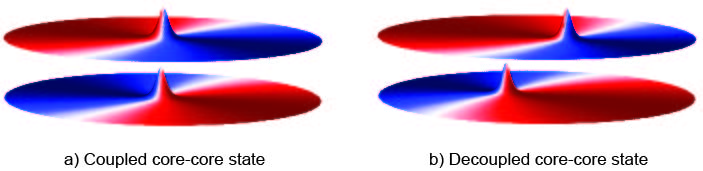}
\caption{Micromagnetically simulated spin distribution showing the coupled (a) and decoupled (b) core-core states of a P-AP vortex pair. The out-of-plane hight reflects the z-component of the core magnetization in the two Py layers, while the blue and red colors correspond to the positive and negative orientation of the in-plane easy-axis component of the magnetization in the vortex periphery. The illustrations are not to scale: the actual separation between the layers in the measured samples is 1 nm or about 1/400th of the lateral particle dimension, such that the spacer would be invisible on the scale shown. The cores were separated by a static field applied in the plane, having the effects of pulling the cores apart in the direction perpendicular to the field since the two chiralities are antiparallel: shown with red-blue versus blue-red for the two vortices in the pair.}
\label{Fig4}
\end{figure*}

RF excitations in the form of shorter pulse-envelopes were sine waves from one period in duration to close to $10^{7}$ periods (about 600 ps to 600 $\mu$s). The measurements were performed at fixed frequency and amplitude with varying number of periods for determining the charteristic core-core switching time and its spread, such as shown in the inset to Fig.2 and Fig.4c,d. No reflection compensation was used, as frequency was not swept.

\section{Micromagnetic simulations}
The micromagnetic simulations were performed using the mumax3 simulation package\footnote{A. Vansteenkiste, et al. AIP Advances 4, 107133 (2014).}. The cellsize was \{x,y,z\}=\{1.76471,1.76471,2\} nm, with 240$\times$200 cells in the x-y plane. The spacer was modeled as a single-cell-thick vacuum layer between the two permalloy disks, each 2 cells in height. The material parameters used were the standard permalloy parameters, $M_s \approx$ 8$\cdot 10^5$ A/m, $A=$13$\cdot 10^{-12}$ J/m, and $\alpha=$ 0.013, with no intrinsic anisotropy. The simulations did not include thermal fluctuations. The simulated spin distributions of the coupled and decoupled core-core states are shown in Fig.\ref{Fig4}.

A set of micromagnetic trajectories under continuous ac field excitation, with the applied biasing dc field corresponding to the center of hysteresis (the mid-point of bi-stability in the core-core potential) are shown in Fig.3c in the main article. The micromagnetically simulated trajectories have qualitatively the same form and evolution as those obtained in the analytical model, shown in Fig.3b. An observation is that the naturally more precise micromagnetic core-core potential is somewhat steeper than the one used in the model and leads to effectively slightly smaller separations between the cores, seen in the micromagnetic trajectories as being more localized near the equilibrium point (tighter trajectory spread in Fig.3c as compared to that in Fig.3b of the main article). Qualitatively, however, our analytics and micromagnetics agree extremely well as regards to the presence of the chaotic dynamics and the period-doubling cascade (Fig.3 of main text).

\section{Theory}

The decoupling process was analyzed in full detail using an analytical model based on the Thiele equations, since investigating the full parameter space using micromagnetic simulations is cumbersome for large systems such as ours, requiring high spatial (sub-nm) and temporal (ps-range) resolution for capturing the core motion. In our theoretical approach, the Thiele equations are complemented with the appropriate core-core interaction potential. The equations of motion can be derived from the Lagrangian, 
\begin{equation}
\mathcal{L} = G \sum_i X_i \dot{Y}_i - U,
\label{eq:lagrangian}
\end{equation}
where $G=\mu_0 L_z M_s/2\gamma$ is the gyroconstant and $\mathbf{X}_i$ are the in-plane displacements of the vortex cores from the center of the particle. The resulting equations of motion are four first-order equations, which can be rewritten in terms of the core-core separation, $\mathbf{x}=(\mathbf{X}_1-\mathbf{X}_2)$, and the pair's 'center-of-mass' offset, $\mathbf{X}=(\mathbf{X}_1+\mathbf{X}_2)/2$. The phase-space of the system is spanned entirely by these four coordinates and is independent of the initial velocities of the two cores.

The potential $U$ describes the forces acting on the individual cores as well as the direct interaction between the cores. An individual core experiences a restoring force from the particle boundary, centering it within the particle. This boundary-restoring force is given by the potential,
\begin{equation}
U_{ms} = \frac{k}{2}\mathbf{X}_i^2 + \frac{k'}{4}\mathbf{X}_i^4,
\end{equation}
where $k=20\mu_0M_s^2L_z^2/9L_x$ and $k' \approx k/2L_x^2$, as shown in \footnote{K.  Yu.  Guslienko,  et al. Journal of Applied Physics, 91, 8037 (2002)}. The force is present due to an additional stray field when the core is displaced from the particle's center.

In stacked vortex pairs, where the vertical core-core spacing is small compared to the core size, the dominant interaction is the direct magnetic dipolar core-core coupling. The inter-vortex interaction through the boundary stray-fields can then be considered negligibly small. The resulting core-core interaction potential is a sum of the four pair-wise interactions among the four poles of the two cores and can be written as
\begin{eqnarray}
\begin{aligned}
U_{cc}(x) = \sigma \mu_0 M_s^2 \Delta^2 \left[ -\Phi \left( \frac{x}{\Delta}, \frac{D}{\Delta}\right)\right.\\ \left. +2\Phi \left( \frac{x}{\Delta}, \frac{D+L_z}{\Delta}\right)-\Phi \left( \frac{x}{\Delta}, \frac{D+2L_z}{\Delta}\right) \right],
\end{aligned}
\end{eqnarray}
where, again, $x$ is the lateral core-core separation, $\Delta$ -- the core size, and $D$ -- the spacer thickness. Function
\begin{eqnarray}
\begin{aligned}
\Phi(x,y) &=\\ &\frac{\pi}{4} \sqrt{2}e^{-x^2/2} \int_0^{\infty} \frac{r dr}{\sqrt{r^2+y^2/2}} e^{-r^2} I_0 (x\sqrt{2}r)
\end{aligned}
\end{eqnarray}
is the universal function describing the normalized potential between the two inner magnetic surfaces and $I_0$ -- the modified Bessel function, obtained assuming a Gaussian distribution of the magnetization  in the core. Parameter $\sigma=\pm 1$ is determined by the relative core polarities and, for the parallel case ($\sigma=1$), with the vertical core-core spacing much smaller than the core length, the interaction is a highly-localized, quasi-monopole core-core attraction. The interaction becomes repulsive when the in-plane core separation is increased to more than a few core radii.

In addition to the core-boundary and core-core interactions, an externally applied field interacts with the in-plane spins in the vortex periphery, outside the core region, which results in a Zeeman force   on the core directed perpendicular to the field. The Zeeman potential is given by
\begin{equation}
U_{Z} = c \chi  \left[\mathbf{e}_z\times\mathbf{X}_i\right]\mathbf{H},
\end{equation}
where the proportionality constant, $c=\pi \mu_0 M_s L_z/2$, is derived using the rigid vortex model and determines the magnitude of the field-induced core movement, while the vortex chirality, $\chi=\pm 1$, determines its the direction.

\begin{figure}[!t]
\includegraphics[width=3in]{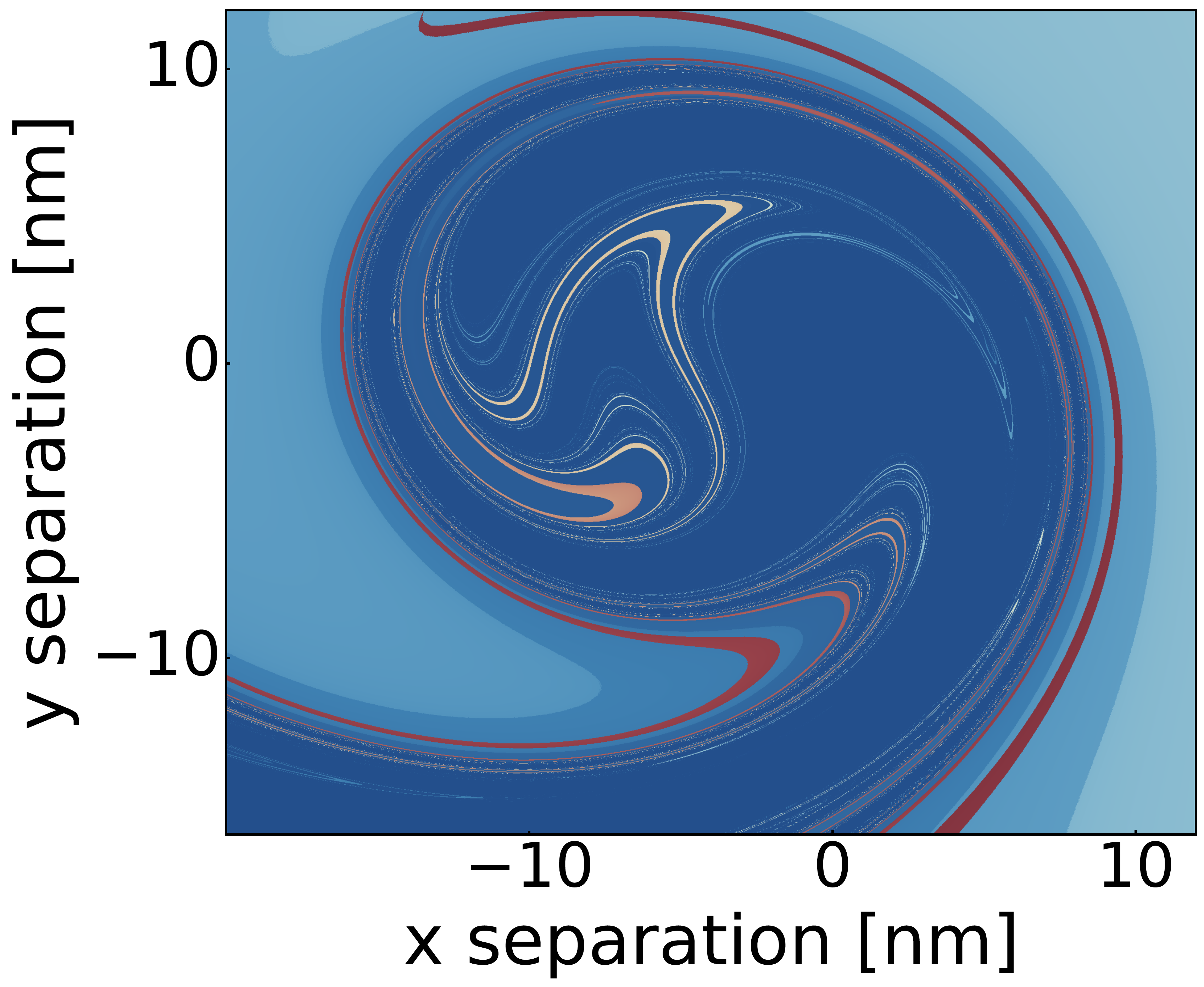}
\caption{Calculated using the developed analytical model basins of attraction for the decoupled state near the coupled state's potential well, for a specific point in time. The color denotes the average number of ac field periods needed for switching (core-core decoupling), with deep blue being infinite and red being 3 periods. The regions of fast switching (few periods) show a fractal behavior (same pattern on zoom in) and correspond to chaotic trajectories within the potential well.}
\label{Fig5}
\end{figure}

The dynamics of the P-AP vortex pair is taken to depend on the core-core separation only since excitation of the 'center-of-mass' motion is forbidden by the symmetry between the two particles and can be excited only at negligible levels through the weak non-linearity of the boundary force, $k'$. The equations of motion for the separation vector are
\begin{eqnarray}
[\mathbf{e}_z \times \mathbf{\dot{x}}]=\omega(|\mathbf{x}|)\mathbf{x}+\lambda \mathbf{\dot{x}}+c\chi(\mathbf{H}_{bias}+\mathbf{h}(t))
\label{eqnarray}
\end{eqnarray}
where $\omega=(\partial U/\partial x)/2Gx$ is the intrinsic oscillation frequency and $\lambda=\pi \alpha \textnormal{ln} (L_x/\Delta)$ -- the geometry-modified effective damping constant. For a time dependent external field the full phase-space becomes three dimensional. Within it, the basins of attractions of the coupled and decoupled core-core states can be mapped out for a resonant excitation, without thermal fluctuations using (\ref{eqnarray}). With the excitation amplitude in the chaotic regime, the two basins near the coupled well are shown in Fig.\ref{Fig5} for a single time-slice.

A study of the effects of thermal fluctuations on the vortex-pair dynamics can be done within a linearized model, whose Green's function is given by 
\begin{equation}
\mathbf{G}(t) = \Theta(t) \begin{pmatrix}\cos \theta t& - \sin \theta t \\ \sin \theta t&\cos \theta t\end{pmatrix}e^{-\lambda \theta t},
\end{equation}
where $\theta = k_l/ G$, $k_l$ -- some linearization parameter, and $\Theta(t)$ -- the Heaviside step function. The time evolution under the influence of a random force, $\mathbf{F}_st(t)$, is then given by
\begin{align}
\mathbf{x}(t) &= \left[x_0 \begin{pmatrix}\cos \theta t \\ -\sin \theta t \end{pmatrix} + y_0 \begin{pmatrix} \sin \theta t \\ \cos \theta t \end{pmatrix} \right]e^{-\lambda \theta t}+\notag\\
&+\int_0^t d\tau ~ \mathbf{G}(t-\tau) \cdot \mathbf{F}_{st}(\tau)
\label{eomlin}
\end{align}
where the random force is assumed to be of white-noise type: $\left\langle F_{i,st}(t)F_{j,st}(t') \right\rangle = \Gamma \delta_{ij} \delta(t-t')$.

\begin{figure}[!t]
\includegraphics[width=3in]{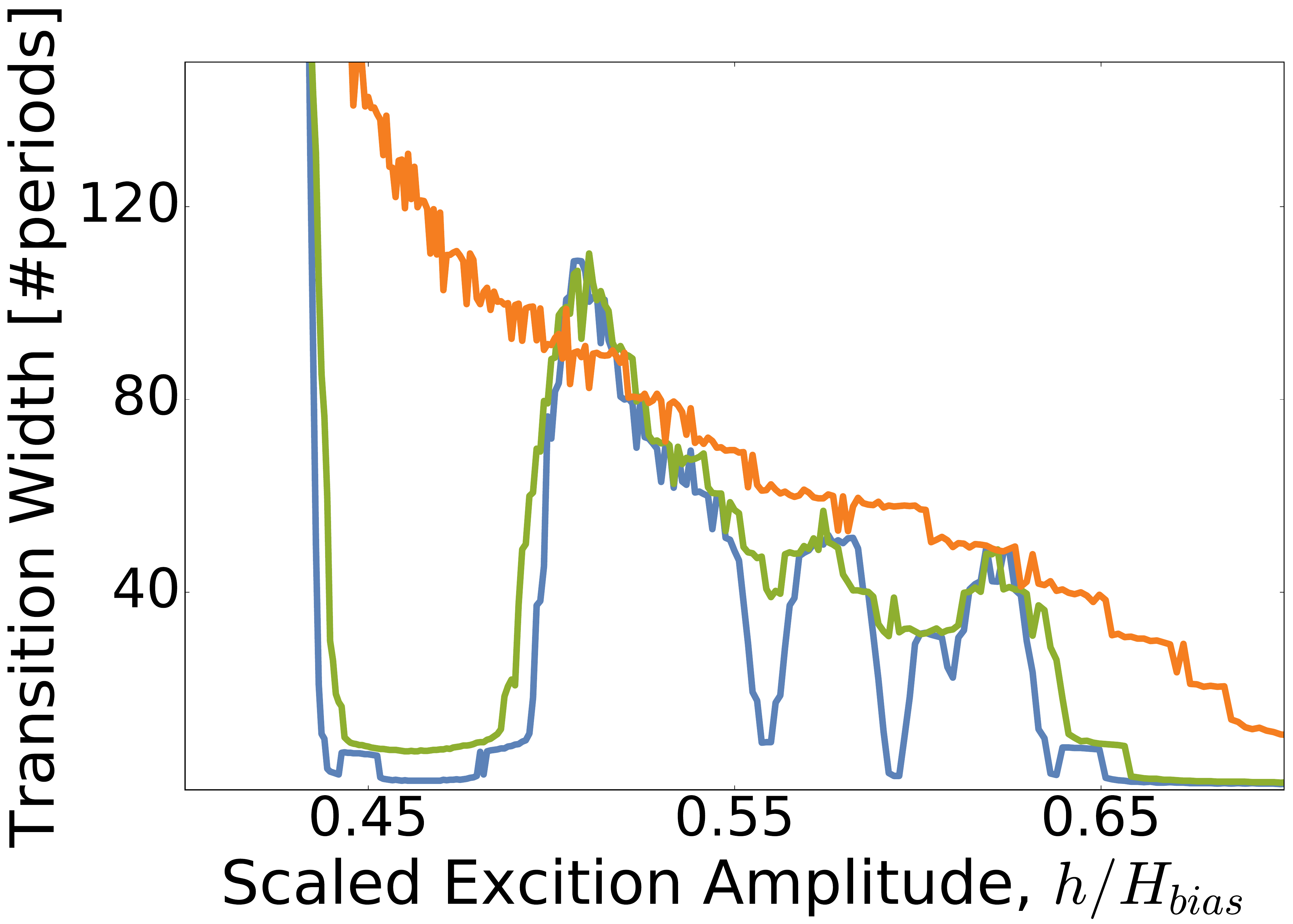}
\caption{Simulated decoupling transition width as a function of normalized ac field amplitude at a fixed frequency of 2.1 GHz. The transition width is taken to be the difference between the number of periods corresponding to the 90th and 10th switching times percentiles. The curves correspond to 80 K (blue; temperature in experiment), 160 K (green) and 240 K (orange). The minima correspond to the chaotic regime. As can be seen, the effect of chaos is suppressed at higher temperatures as well as the hysteresis itself (fully confirming the experiment).}
\label{Fig6}
\end{figure}

Next, $B_{ij}(t '-t) = \left\langle x_i(t)x_j(t') \right\rangle$ is calculated:
\begin{equation}
\mathbf{B}(t) = \frac{\Gamma}{2\lambda \omega}  \begin{pmatrix}\cos \theta t& - \sin \theta t \\ \sin \theta t&\cos \theta t\end{pmatrix} e^{-\lambda \theta |t|}.
\end{equation}
For the special case of $t=t'$ it can be compared to the thermodynamic result:
\begin{equation}
\left\langle x_i^2 \right\rangle = \frac{k_B T}{k_l}.
\end{equation}
This gives the magnitude of the random force,
\begin{equation}
\Gamma = \frac{2\lambda k_B T}{G},
\label{fstmag}
\end{equation}
which does not depend on the linearization parameter, $k_l$. For core motion offset from the particle center the system can still be linearized to the same form as (\ref{eomlin}), since (\ref{fstmag}) does not depend on the linearization parameter. The average decoupling time under the influence of an external ac field with the thermal fluctuations included was calculated. We find that the width of the transition from almost no decoupling to almost full decoupling, at some resonant frequency, shows a non-monotonous behavior versus the ac field amplitude. As shown in the main text by comparing the theory versus experiment in Fig.4(b) vs Fig.4(d), this behavior is due to chaotic dynamics in the system. Fig.\ref{Fig6} shows how the core-core decoupling transition width goes from highly non-monotonous to clearly monotonous at higher temperatures, where the effects of chaos are suppressed. Our experiments at room temperature fully confirm this thermal transition -- the core-core decoupling at room temperature is fully stochastic with no core-core bi-stability or hysteresis [Fig.1(b) of main text].

The Lyapunov characteristic exponents (LCE) of a system characterize the stability of a given trajectory to small fluctuations. The time dependent deviation, $\delta \mathbf{x}(t)$, from a trajectory evolves, in the linear approximation, as
\begin{eqnarray}
\delta\mathbf{x}(t) = |\delta \mathbf{x}_0| e^{\lambda_{LCE} t}
\end{eqnarray}
from some infinitesimal initial deviation, $|\delta \mathbf{x}_0|$, with the Lyaponov characteristic exponents $\lambda_{LCE}$ (one LCE per dimension of phase-space). Positive LCE indicate an unstable motion, for which the deviation grows with time, while negative LCE characterize stable motion. Generally, in three dimensions, a chaotic motion is characterized by a set of one positive, one negative, and one zero-valued LCE.

The Lyapunov characteristic exponents for our system, shown in Fig.4(b) of the main article, were calculated using the methods described in \footnote{G. Benettin, et al. Meccanica 15, 9 (1980).}. Our algorithm is based on \footnote{M. Sandri, The Mathematica Journal, 6, 3, p.78-84 (1996)}, implemented in Mathematica, with our code available at \footnote{https://github.com/Artemkth/Mathematica-LCE-calculator}.


